\documentclass[journal=nalefd, manuscript=article]{achemso}
\setkeys{acs}{email = false}

\usepackage[version=3]{mhchem} 
\usepackage{amsmath}
\usepackage{amssymb}
\usepackage{amsfonts}
\usepackage[overload]{empheq}
\usepackage[normalem]{ulem}
\usepackage{graphicx} 
\graphicspath{ {./Figs/} }
\usepackage[utf8]{inputenc}
\usepackage[english]{babel}
\usepackage[colorinlistoftodos]{todonotes}
\usepackage{mathtools}
\usepackage{float}
\usepackage{array,multirow}
\usepackage{bm}
\usepackage{cases}
\usepackage[position=top]{subfig}
\usepackage{etoolbox}
\usepackage{pdfpages}
\newcommand{\mhat}{\hat{\mathbf{m}}}
\newcommand{\Hvec}{\mathbf{H}}

\title{Anisotropic Gigahertz Antiferromagnetic Resonances of the Easy-Axis van der Waals Antiferromagnet CrSBr}
\begin{document}
\noindent\large{Thow Min Jerald Cham\textsuperscript{a$\ddagger$}, Saba Karimeddiny\textsuperscript{a,\#}, Avalon H. Dismukes\textsuperscript{b}, Xavier Roy\textsuperscript{b}, Daniel C. Ralph\textsuperscript{ac*}, Yunqiu Kelly Luo\textsuperscript{acd$\ddagger$*}}\\
\small{\noindent\textit{\textsuperscript{a} Cornell University, Ithaca, NY 14850, USA}\\
\textit{\textsuperscript{b} Department of Chemistry, Columbia University, New York, NY 10027, USA}\\
\textit{\textsuperscript{c} Kavli Institute at Cornell, Ithaca, NY 14853, USA}\\
\textit{\textsuperscript{d} Department of Physics and Astronomy, University of Southern California, Los Angeles, CA 90089, USA}\\
\textsuperscript{*}Email: dcr14@cornell.edu\\
\textsuperscript{*}Email: yl664@cornell.edu, kelly.y.luo@usc.edu\\
\textit{\textsuperscript{$\ddagger$} T.M.J.C and Y.K.L contributed equally to this paper}}

\begin{abstract}
We report measurements of antiferromagnetic resonances in the van der Waals easy-axis antiferromagnet CrSBr. The interlayer exchange field and magnetocrystalline anisotropy fields are comparable to laboratory magnetic fields, allowing a rich variety of gigahertz-frequency dynamical modes to be accessed. By mapping the resonance frequencies as a function of the magnitude and angle of applied magnetic field we identify the different regimes of antiferromagnetic dynamics. The spectra show good agreement with a Landau-Lifshitz model for two antiferromagnetically-coupled sublattices, accounting for inter-layer exchange and triaxial magnetic anisotropy. Fits allow us to quantify the parameters governing the magnetic dynamics: at 5 K, the interlayer exchange field is $\mu_0 H_E =$ 0.395(2) T, and the hard and intermediate-axis anisotropy parameters are $\mu_0 H_c =$ 1.30(2) T and $\mu_0 H_a =$ 0.383(7) T. The existence of within-plane anisotropy makes it possible to control the degree of hybridization between the antiferromagnetic resonances using an in-plane magnetic field.

\end{abstract}

van der Waals (vdW) antiferromagnets (AFs), a new class of exfoliatable magnetic materials, provide opportunities for future memory, logic, and communications devices because their magnetic properties are highly tunable, for example by applied electric fields \cite{Jiang2018, Zhang2020} or strain \cite{Cenker2022}, they can possess long-lived magnons \cite{Bae2022}, and they can be straightforwardly integrated within complex heterostructures by mechanical assembly \cite{zhong2017, Geim2013}. The strength of exchange coupling between vdW  layers is typically much weaker than the direct exchange or superexchange coupling in ordinary 3D-crystal antiferromagnets. This weak interlayer exchange yields antiferromagnetic resonances in the gigahertz range, rather than the terahertz range that is more typical for antiferromagnets \cite{Johnson1959, Sievers1963, Kondoh1970, Kampfrath2011, Baierl2016, Li2020, Vaidya2020}. The resonance modes in vdW antiferromagnets can therefore be addressed and controlled with conventional microwave electronics.

\vspace{0.5 cm}Here, we use antiferromagnetic resonance measurements to map the gigahertz-frequency resonances in the vdW antiferromagnet CrSBr \cite{Telford2020,  Lee2021, Wilson2021, Telford2022,Bae2022}. Unlike the only other two vdW antiferromagnets whose modes have been characterized in detail previously (CrI$_3$ \cite{Zhang2020, shen2021multi} and CrCl$_3$ \cite{Macneill2019}), CrSBr has a significant triaxial magnetic anisotropy with an easy axis oriented within the vdW plane \cite{Yang2021}. This anisotropy modifies the form of the antiferromagnetic resonances, and for a sufficiently strong magnetic field aligned with the anisotropy axis it induces a discontinuous transition that changes the mode spectrum abruptly. By measuring the resonances in CrSBr as a function of the angle and magnitude of applied magnetic field and comparing to a Landau-Lifshitz (L-L) model, we identify the resonance modes in the frequency range 1 - 40 GHz, make a quantitative determination of the interlayer exchange and anisotropy fields, and show that the hybridization between modes can be controlled with an in-plane magnetic field.  These measurements provide the foundational understanding needed for efforts to utilize these modes within future spintronic devices.

\vspace{0.5 cm}
 CrSBr is an A-type antiferromagnetic vdW semiconductor with intralayer ferromagnetic coupling and interlayer antiferromagnetic coupling  \cite{Telford2020,Yang2021,Telford2022}.  It has a bulk N\'eel temperature ($T_\text{N}$) of 132 K and an intermediate ferromagnetic phase with Curie temperature ($T_\text{C}$) in the range of 164 - 185 K as measured using transport and optical methods \cite{Telford2020, Lee2021, Telford2022}. Each vdW layer consists of two buckled rectangular planes of Cr and S atoms sandwiched between Br atoms. The layers are stacked along the \emph{c}-axis through vdW interactions to form an orthorhombic structure (space group \emph{Pmmn}) (Fig.\ 1a and Fig.\ 1b). In our measurements, we use millimeter-length, flat, needle-like single crystals of CrSBr grown by a modified chemical vapor transport approach (see growth details in the Methods section). The long axis of the needle is oriented along the crystallographic \emph{a}-axis and the flat face corresponds to the vdW plane (supplementary information, Fig.\ S1). Previous magnetoresistance studies identified a magnetic easy axis (N\'eel axis) along the crystallographic \emph{b}-axis (blue and green arrows in Fig.\ 1b and 1c), an intermediate axis along the \emph{a}-axis, and a hard axis along the \emph{c}-axis \cite{Telford2020,Telford2022}.

\begin{figure}[htpb]
    \centering
    \includegraphics[width=0.96\textwidth]{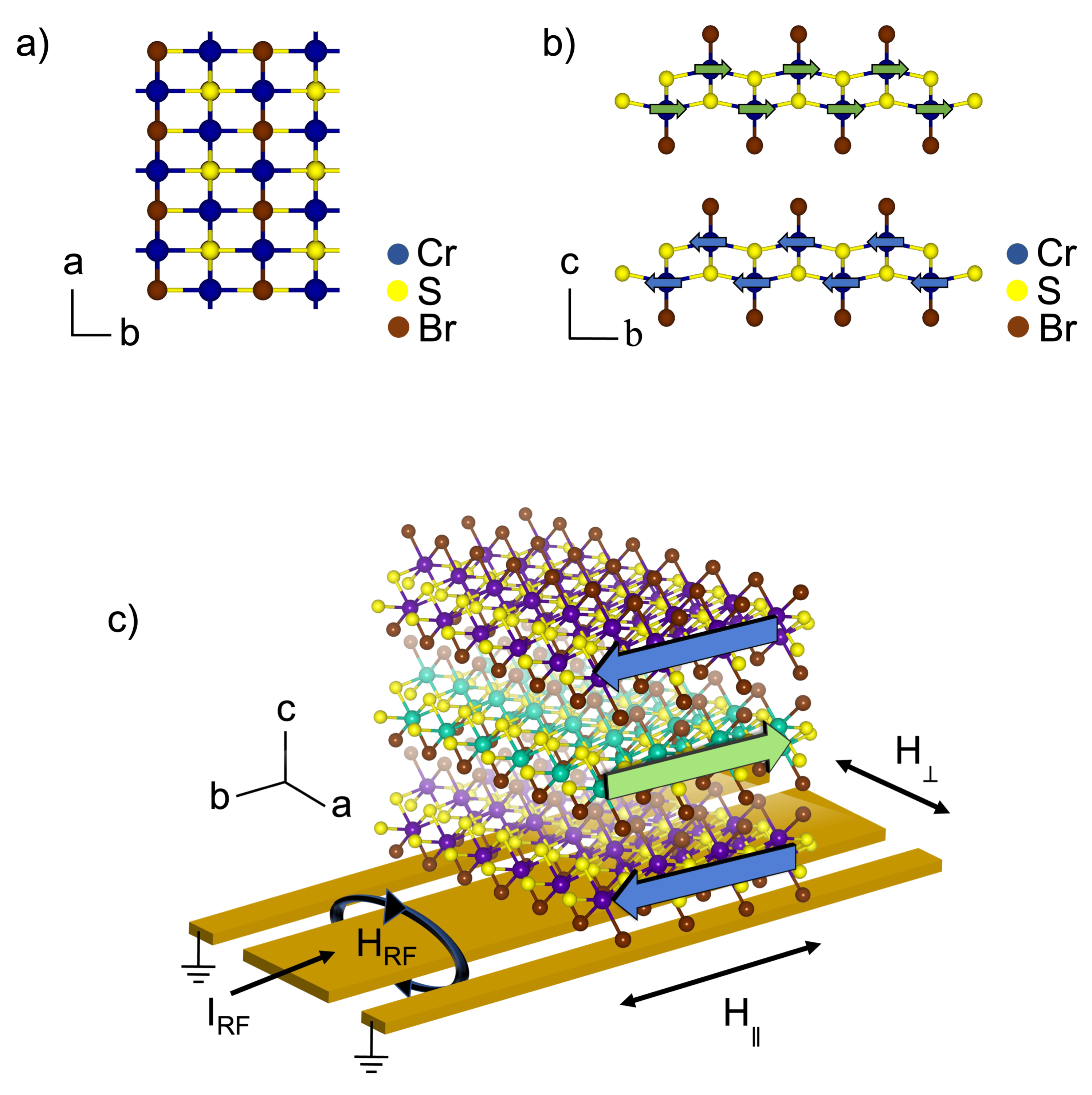}
    \label{Figure1}
\caption{\textbf{Crystal structure and experimental setup for the coplanar waveguide measurements.} (a) Crystal structure of CrSBr as viewed along the out-of-plane $c$-axis, with Cr, S and Br atoms represented by blue, yellow and red spheres respectively. (b) Crystal structure of two CrSBr vdW layers as viewed along the $a$-axis. Blue and green arrows indicate the antiferromagnetic in-plane magnetic order with the N\'eel vector $\hat{N}$ aligned with the easy magnetic axis, which is the crystal $b$-axis. (c) Experimental schematic of a CrSBr crystal mounted on a coplanar waveguide for microwave absorption measurements (not to scale) \cite{Momma}. Here, we show just 3 vdW layers of the bulk crystal for clarity. The long axis of the crystal ($a$-axis) is aligned perpendicular to the signal line such that the microwave magnetic fields are perpendicular to the magnetic easy axis. The purple and green balls represent Cr atoms in adjacent layers, with magnetization polarized to the left (blue arrows) or right (green arrow) indicating the N\'eel order. External fields can be applied either parallel ($H_{\parallel}$) or perpendicular ($H_{\perp}$) to $\hat{N}$.}
\end{figure}

\vspace{0.5 cm}
We investigate the antiferromagnetic resonances of CrSBr by placing the crystal on a coplanar waveguide and measuring the microwave absorption spectrum using a two-port vector network analyzer. For the measurements reported in the main text, we align the crystal with its long axis (i.e., the \emph{a} axis) perpendicular to the waveguide, such that the N\'eel axis $\hat{N}$ (i.e., the \emph{b} axis) is perpendicular to the in-plane RF field ($H_{RF}$). An external DC magnetic field is swept in-plane, either perpendicular ($H_\perp$), parallel ($H_\parallel$), or at intermediate angles relative to $\hat{N}$ (Fig.\ 1c). (See supplementary information, Section I for spectrum acquisition details.) We use a low microwave power, -20 dBm, to avoid any significant sample heating and to measure the magnetic resonance induced microwave absorption in the linear response regime. We verify that the spectra are insensitive to the incident power in this regime (supplementary information, section VIII). Representative transmission spectra as a function of applied field at 5 K are shown in Fig.\ 2. Dark green features represent strong microwave absorption due to resonance modes.  We observe two resonance modes in the $H_\perp$ configuration and show their dependence on magnetic field in Fig.\ 2a up to the maximum applied field of $\pm$ 0.57 T. Below, we will identify the two resonances in Fig.\ 2a as acoustic and optical modes originating from an initial spin-flop configuration in which the two  spin sublattices are canted  away from the easy axis. In the $H_\parallel$ case (Fig.\ 2b), two resonance features are also observed, but with opposite signs of concavity ($d^2f/dH^2$) compared to Fig.\ 2a.  At approximately $H_\parallel = \pm$ 0.4 T in Fig.\ 2b there is a sudden transition beyond which only one mode is observed. 
A previous  study associated a related transport signal as due to a spin-flip transition, in which the external field overcomes the anisotropy field to abruptly align the two spin sublattices from an antiparallel to parallel orientation \cite{GOSER1990}.  We will find that the anisotropy parameter $H_a$ is slightly smaller than the interlayer exchange field $H_E$, so strictly speaking the discontinuity is likely a spin-flop transition in which the 
N\'eel vector initially reorients by 90$^\circ$ and the spin-sublattices tilt to form a canted state within a narrow window of applied magnetic field, before eventually they saturate
to a parallel alignment as a function of increasing field
\cite{Baltz2018}.

\begin{figure}[htpb]
    \centering
    \includegraphics[width=1.0\textwidth]{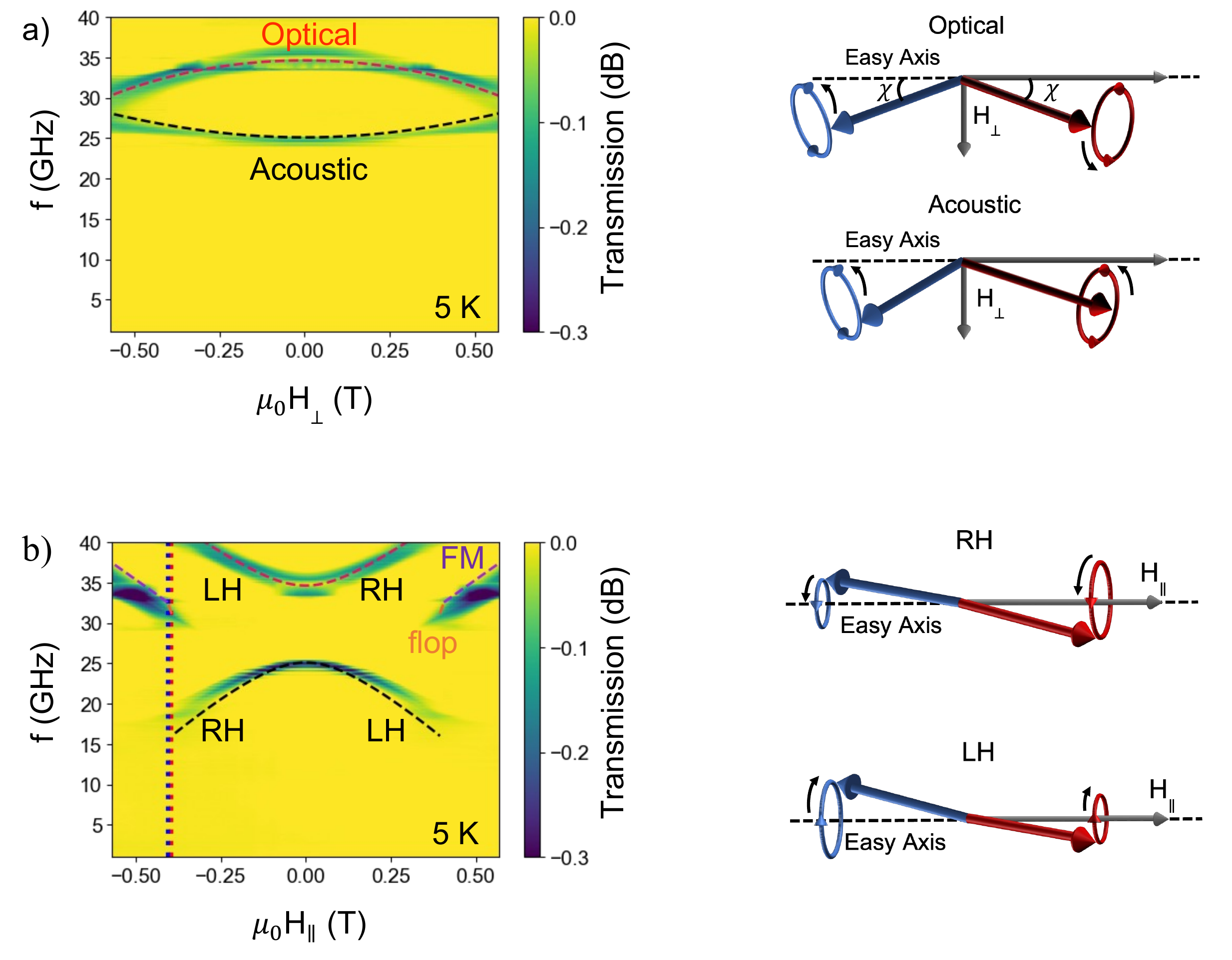}
    \label{fig:5K}
    \caption{\textbf{Microwave absorption spectra measured at 5 K as a function of magnetic field applied perpendicular ($\boldsymbol{H_{\perp}}$) or parallel ($\boldsymbol{H_{\parallel}}$) to $\boldsymbol{\hat{N}}$.} 
(a) Microwave transmission (S$_{21}$) signal as a function of $H_{\perp}$,  magnetic field applied along the crystal $a$ axis. (b) The corresponding spectra as a function of $H_{\parallel}$, magnetic field applied along the crystal $b$ axis. S$_{21}$ values are shown relative to a field-independent subtracted background. Dashed lines show a fit to the results of the L-L model (Eqs.\ (\ref{eqn:omega1perp})-(\ref{eqn:FM})). Diagrams on the right illustrate the form of some of the resonant modes.}
\end{figure}

\vspace{0.5 cm}
We can identify the nature of the detected resonance modes by modeling them with two coupled L-L equations using a macrospin approximation for each spin sublattice.  This is a reasonable approximation since the applied DC and AC magnetic fields, as well as the interlayer exchange, are small perturbations relative to the  intralayer ferromagnetic coupling \cite{Yang2021}, so they should not affect the net magnetization of individual atomic layers significantly. Denoting the magnetic-moment direction of the two spin sublattices with the unit vectors $\mhat_{1}$ and $\mhat_{2}$, we model the interlayer Weiss exchange field acting on $\mhat_{1(2)}$ as  $-H_E\mhat_{2(1)}$. The triaxial magnetic anisotropy can be modelled by including the hard axis and intermediate axis anisotropy fields   $-H_c (\mhat_{1(2)} \cdot \hat{c})\hat{c}$ and $-H_a (\mhat_{1(2)} \cdot \hat{a})\hat{a}$, where $H_c$ and $H_a$ are constants and $\hat{c}$ and $\hat{a}$ are unit vectors along the crystal's \emph{c} and \emph{a} axes. The coupled two-lattice L-L equations can then be written as \cite{Keffer1952}
\begin{align}
    \begin{split}
        \frac{d\mhat_{1(2)}}{dt} = &-\mu_{0}\gamma \mhat_{1(2)} \times ( \Hvec - H_E \mhat_{2(1)} - H_c(\mhat_{1(2)} \cdot \hat{c})\hat{c} - H_a (\mhat_{1(2)} \cdot \hat{a})\hat{a}).
    \end{split}
\label{eqn:LL}
\end{align}
 Here $\mu_0$ is the permeability constant, $\gamma$ is the gyromagnetic ratio, and $\Hvec$ is the externally applied magnetic field.  Since we will focus only on the relationship between the resonance frequency and applied magnetic field, but not the resonance linewidth, we omit a damping term.

\vspace{0.5cm}
In the $H_\perp$ configuration, since the external field is  perpendicular to the N\'eel axis the initial spin configuration (before $H_{RF}$ is applied) is a spin-flop state in which the two spin sublattices rotate symmetrically towards $H_\perp$  as illustrated in Fig.\ 2a, with a tilt angle $\sin\chi = H_\perp/(2H_E+H_a)$. Upon excitation by $H_{RF}$, assuming small damping, the magnetic sub-lattices will oscillate at frequencies corresponding to the normal modes of the system. Taking terms to the first order in the excitation amplitudes $\delta \mhat_{1(2)}$, we calculate these frequencies by rewriting Eq.\  (\ref{eqn:LL}) as a 6 $\times$ 6 matrix in the $\delta m^{1(2)}_{b,a,c}$ basis. Solving for the eigenvalues, and taking only the non-trivial positive solutions, we obtain the following resonance frequencies (details in supplementary information, section II). 
\begin{align}
        \omega_{1,\perp} &= \mu_0 \gamma \sqrt{\frac{(H_\perp^{2}(2H_E - H_a) + H_a(H_a + 2H_E)^2)(2H_E + H_c)}{(H_a + 2H_E)^2}}
\label{eqn:omega1perp}
\end{align}
\begin{align}
    \omega_{2,\perp} &= \mu_0 \gamma \sqrt{\frac{((H_a + 2H_E)^2 - H_\perp^2)H_c}{H_a + 2H_E}}
\label{eqn:omega2perp}
\end{align}
These equations should apply up to the field $|H_\perp| = 2H_E + H_a$ at which point the applied field drives the spin sublattices fully parallel, which is a field beyond the range of our measurements.  When $\mu_0H_a = 0$ T, these modes simplify to the form of acoustic ($\omega_{1,\perp}$) and optical ($\omega_{2,\perp}$) modes previously reported for easy-plane antiferromagnets \cite{Streit1980, Macneill2019}. When $\mu_0H_a \neq 0$ T, the acoustic mode has a finite frequency as $H_\perp$ goes to zero, instead of being proportional to $H_\perp$.  Even when $\mu_0H_a \neq 0$ T, the sample in this field geometry is symmetric under twofold rotation around the applied field direction combined with sublattice exchange \cite{Macneill2019}.  The acoustic and optical modes have opposite parity under this two-fold rotation, so they remain unhybridized and they cross when they intersect (see the 50 K and 100 K data in Fig.\ 3a,c), rather than undergoing an avoided crossing (supplementary information, section IV).

\vspace{0.5cm}
In the $H_{\parallel}$ configuration, the applied  field is parallel to the N\'eel axis and for small fields ($|H_\parallel| < \sqrt{(2H_E-H_a)H_a}$ if $H_a < H_E$, or $|H_\parallel| < H_E$ if $H_a > H_E$) the spin sub-lattices initially stay antiparallel to each other and aligned along this preferred axis. By a matrix diagonalization procedure similar to the one discussed above, for this antiparallel configuration we obtain the resonance frequencies
\begin{align}
        \omega_{1,\parallel}&= \mu_0 \gamma \sqrt{H_\parallel^2 + H_a(H_E + H_c) + H_EH_c - \sqrt{H_\parallel^2(H_a + H_c)(H_a + 4H_E + H_c) + H_E^2(H_a - H_c)^2}}
\label{eqn:omega1para}
\end{align}
\begin{align}
    \omega_{2,\parallel} = \mu_0 \gamma \sqrt{ H_\parallel^2 + H_a(H_E + H_c) + H_EH_c + \sqrt{H_\parallel^2(H_a + H_c)(H_a + 4H_E + H_c) + H_E^2(H_a - H_c)^2} }.
\label{eqn:omega2para}
\end{align}
These modes correspond to linear superpositions of the left-handed (LH) and right-handed (RH) modes previously reported for easy-axis antiferromagnets \cite{Keffer1952, Li2020}. Here, the presence of triaxial anisotropy ($H_c$ $\neq$ $H_a$) induces an avoided crossing at zero field (Fig.\ 2b) and causes the eigenmodes at zero field to be  symmetric and antisymmetric combinations of the LH and RH modes \cite{Liensberger2019}.

\vspace{0.5cm}
If $H_a < H_E$, in this model as a function of increasing $H_\parallel$ the antiferromagnet  undergoes a spin-flop transition at $H_\parallel = \sqrt{(2H_E-H_a)H_a}$ in which case the N\'eel vector abruptly reorients to be perpendicular to the applied field and each spin sublattice cants toward the field direction with a tilt angle $\sin \chi = H_\parallel/(2H_E - H_a)$. For this configuration, we obtain the resonance frequencies
\begin{align}
    \omega_{1,\parallel,\text{flop}} &= \mu_0\gamma \sqrt{\frac{(H_\parallel^2(2H_E + H_a)-H_a(2H_E - H_a)^2)(2H_E+H_c-H_a)}{(2H_E - H_a)^2}}\label{flop_ac}\\
    \omega_{2,\parallel,\text{flop}} &= \mu_0\gamma\sqrt{\frac{((2H_E - H_a)^2-H_\parallel^2)(H_c-H_a)}{2H_E - H_a}}.
    \label{flop_op}
\end{align}
For this case that $H_a<H_E$, as $H_\parallel$ is increased further the canting angle of the spin sublattices in the spin flop state increases continuously, and eventually approaches $\chi = \pi/2$  reaching the fully-aligned spin state for $H_\parallel \ge 2H_E - H_a$.  For the alternative case that $H_a > H_E$, within this model the spin-flop state is never stabilized, and the antiferromagnet makes a discontinuous spin-flip transition directly from the anti-aligned state to the fully-parallel state at $H_\parallel = H_E$.  Once the antiferromagnet is in the fully-parallel state for large $H_\parallel$, in either case the calculated resonance frequencies take the form
\begin{align}
    \omega_{1, \parallel, FM} &= \mu_0 \gamma \sqrt{(H_\parallel+H_a)(H_\parallel+H_c)}
\label{eqn:FM}
\end{align}
\begin{align}
    \omega_{2, \parallel, FM} &= \mu_0\gamma \sqrt{(H_\parallel+H_a-2H_E)(H_\parallel+H_c-2H_E)}.
    \label{eqn:QFM}
\end{align}
However, we do not expect either $\omega_{2,\parallel,\text{flop}}$ or $\omega_{2, \parallel, FM}$ to be detectable in our measurements because these modes are even as a function of 2-fold rotation about the $H_\parallel$ axis, and the oscillating RF field has only odd components for this configuration \cite{Macneill2019}.   

\vspace{0.5cm}
The dashed lines in Fig.\ 2a represent fits of the resonant modes for the $H_{\perp}$ configuration to Eqs.\ (\ref{eqn:omega1perp}) and (\ref{eqn:omega2perp}), and in Fig.\ 2b we present the analogous fits of the modes for the $H_{\parallel}$ configuration to Eqs.\ (\ref{eqn:omega1para})-(\ref{flop_ac}), (\ref{eqn:FM}) taking into account the predicted transition fields. The fits provide a good description of all the observed modes with a common set of fit parameters. The values of the fit parameters for a simultaneous least-squares fit to the data in Figs.\ 2a and 2b are an inter-layer exchange field strength of $\mu_0H_{E}$ = 0.395(2) T and anisotropy field strengths of $\mu_0H_{a}$ = 0.383(7) T along the in-plane intermediate axis and $\mu_0H_{c}$ = 1.30(2) T along the out-of-plane hard axis.  As expected the exchange field is an order of magnitude smaller than the typical exchange in bulk antiferromagnets.  It is larger, however, than the value of 0.1 T measured in CrCl$_3$ using similar techniques \cite{Macneill2019}.  The modes that are most prominent in the spectra correspond to the spin-flop configuration for the $H_\perp$ geometry, the antiparallel spin configuration at low values of $H_\parallel$, and the fully-parallel spin alignment at large values of $H_\parallel$.  However, because by our fits $H_a$ is slightly smaller than $H_E$, we do also anticipate that there could in principle be a very small region corresponding to the spin-flop state in Fig.\ 2b.

\vspace{0.5cm}
Next, we investigate the evolution of the resonant modes with temperature. As shown in Figs.\ 2a,b and Figs.\ 3a-h, we observe qualitatively similar resonance features over the temperature range from 5 to 100 K.  With increasing temperature, the modes shift to lower frequency and the magnetic field scales decrease for both the value of $H_\perp$ where the two modes become degenerate and for the value of $H_\parallel$ corresponding to the discontinuous transition. These observations can be attributed to decreasing values of all of the exchange and anisotropy parameters $H_E$, $H_a$ and $H_c$  with increasing temperature. Figure 4 plots the values of $H_E$, $H_a$, and $H_c$ extracted from simultaneous fits of equations (2)$-$(6) to the $H_{\parallel}$ and $H_{\perp}$ transmission spectra for a series of temperatures from 5 to 128 K. We observe a monotonic decrease in all three parameters.

\begin{figure}[htpb]
    \centering
    \includegraphics[width=0.85\textwidth]{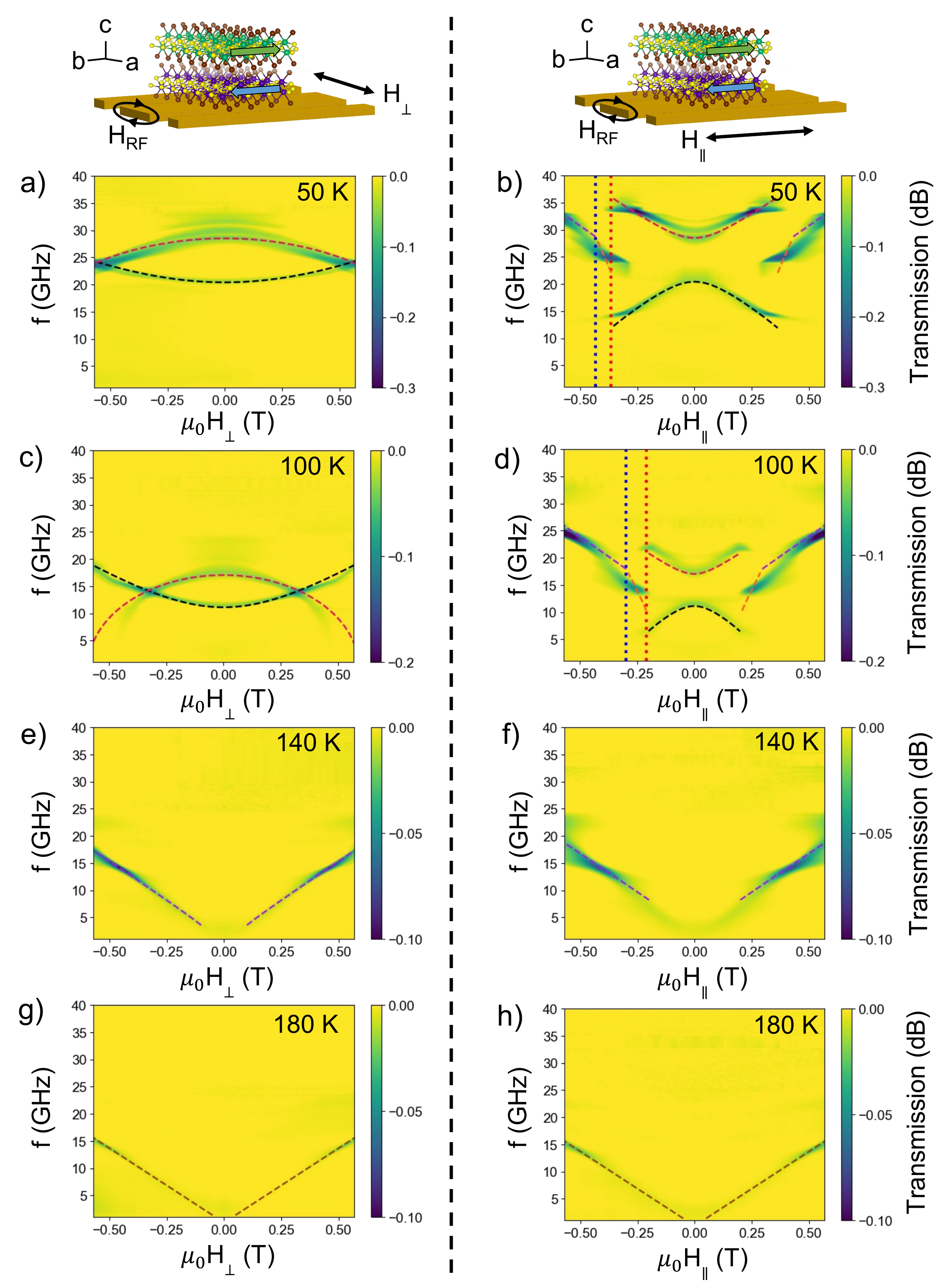}
    \label{Figure3}
    \caption{\textbf{Fitted absorption spectra at different temperatures.} 
(a, c, e, g) Microwave transmission (S$_{21}$) signals as a function of $H_{\perp}$, magnetic field applied along the crystal $a$ axis, for the selected temperatures indicated.  (b, d, f, h) S$_{21}$ signals as a function of $H_{\parallel}$ at the same selected temperatures.  Dashed lines are fits to Eqs.\ (\ref{eqn:omega1perp})-(\ref{eqn:FM}) with parameters $H_E$, $H_a$, and $H_c$ that are allowed to vary with temperature but are independent of the applied field magnitude and direction. Red and blue dotted lines represent the predicted spin-flop and fully-parallel spin alignment transition fields respectively.}
\end{figure}

\begin{figure}[htpb]
    \centering
    \includegraphics[width=0.8\textwidth]{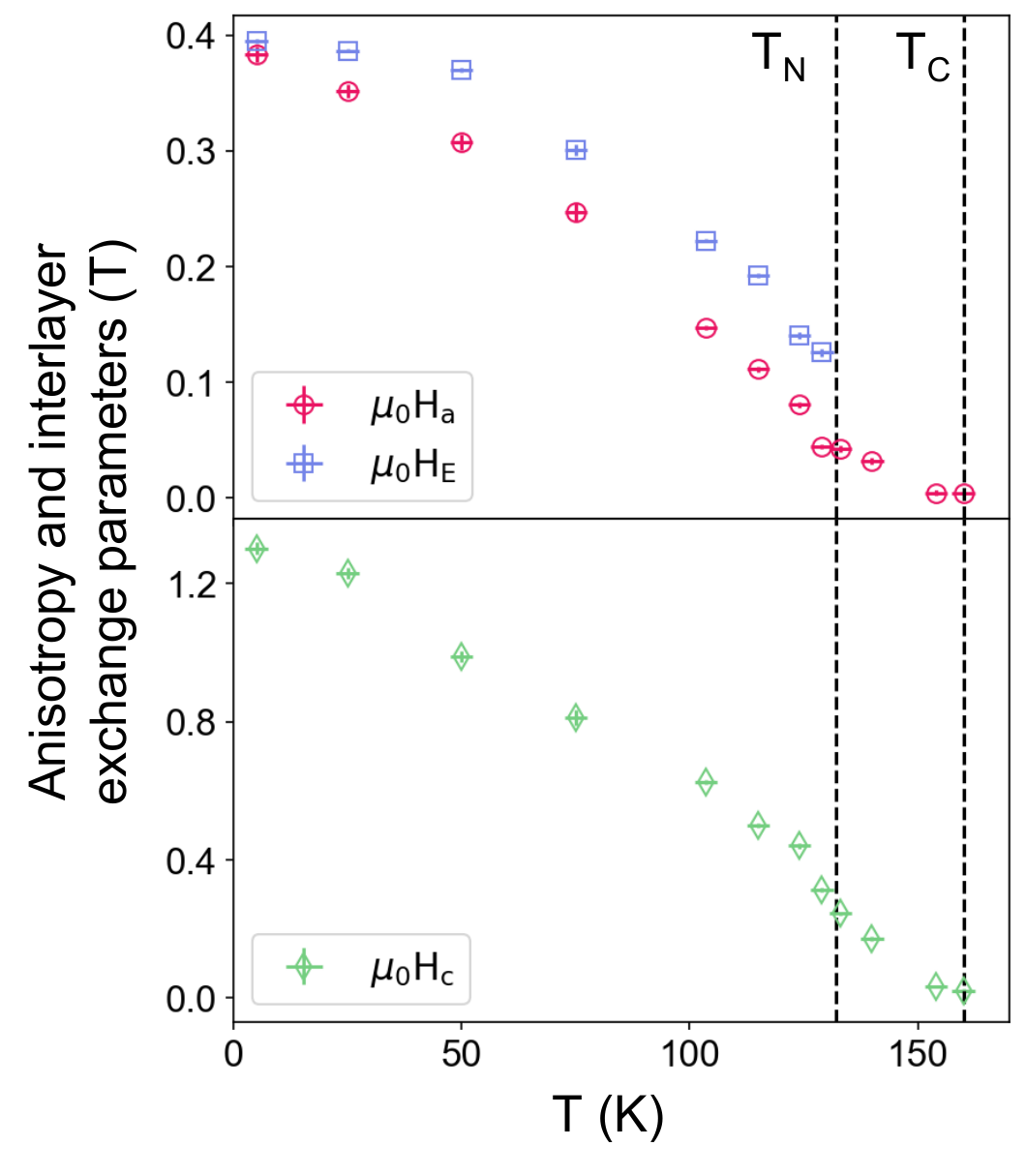}
    \label{Figure4}
    \caption{\textbf{Temperature dependence of the interlayer exchange and the anisotropy field strengths $\boldsymbol{H_{E}}$, $\boldsymbol{H_{a}}$, and $\boldsymbol{H_{c}}$.} a) Temperature dependence of the interlayer exchange $H_{E}$ (blue squares) and the in-plane easy-axis anisotropy parameter $H_{a}$ (red circles). b) Temperature dependence of out-of-plane anisotropy parameter $H_{c}$. The black dashed lines indicate the estimated N\'eel temperature $T_N \approx 132$ K and the Curie temperature $T_c \approx 160$ K, previously measured in magnetometry and magnetotransport measurements \cite{Telford2020, Lee2021, Telford2022}.}
\end{figure}

\vspace{0.5cm}
The anisotropy parameter $H_a$ decreases more quickly with increasing temperature than $H_E$, which within our model should open up a larger window of field within which the spin-flop configuration is stabilized in the $H_\parallel$ geometry.  This regime corresponds to the one feature of our measurement which is not captured well by the fits -- see how the field dependence of the measured frequency deviates from the prediction for the spin-flop state for values of $|H_\parallel|$ just above the discontinuous jumps for the 50 K and 100 K data in Figs.\ 3b and 3d.  We take this as a hint that our simple model may not fully capture the angular dependence of the exchange energy or the magnetic anisotropy for large canting angles within the spin-flop state.

\vspace{0.5cm}
At 140 K (Figs. 3e, f), the resonance spectra become quite different compared to the lower-temperature measurements, with only a single resonance mode observed, rather than two. The frequency dependence can be fit well to a ferromagnetic dependence (Eq.\ (\ref{eqn:FM})) with the in-plane easy-axis anisotropy persisting to give slightly different frequencies for H$_{\parallel}$ vs.\ H$_{\perp}$.  This measurement is performed above the N\'eel temperature of 132 K, so our interpretation is that this spectrum corresponds to the intermediate ferromagnetic state observed previously by transport and optical second harmonic generation \cite{Telford2020, Lee2021,Telford2022}.  Finally, by 180 K (Figs.\ 3g, 3h) the anisotropy is no longer visible and only a linear electron paramagnetic resonance signal remains, with a slope of 26.6(1) GHz/T and a very narrow linewidth \cite{Stanger1997}.

\vspace{0.5cm}
We have explored in more detail the relationship between the $H_{\parallel}$ and $H_{\perp}$ modes by measuring the evolution of the modes when the external field is applied at angles $\phi$ between the $b$ crystal axis (easy anisotropy axis) and the $a$ crystal axis (intermediate anisotropy), where $\phi = 0^\circ$ corresponds to the  $H_\parallel$ configuration. As shown in Fig.\ 5a for data measured at 100 K, the convex-shaped optical and acoustic modes $H_{\perp}$ modes ($\phi$ = 90$^\circ$) flatten out and evolve toward the concave-shaped $H_{\parallel}$ modes as $\phi$ is decreased. As this happens, the simple crossing between the modes present for $\phi = 90^\circ$ evolves into the avoided crossing visible for the 75$^\circ$ and 60$^\circ$ data. This hybridization in the modes as $\phi$ is decreased from 90$^\circ$ is analogous to the opening of an anti-crossing hybridization gap in CrCl$_3$ due to breaking of 2-fold rotational symmetry by an out-of-plane applied magnetic field \cite{Macneill2019}, but here the in-plane easy axis allows for two-fold rotational symmetry to be broken with an in-plane field oriented away from the $H_\perp$ direction (see further analysis in the supplementary information, section IV and VII). As a function of decreasing $\phi$ from 90$^\circ$, the discontinuity corresponding to the spin-flip transition first becomes visible at about $\phi = 60^\circ$ and is increasingly prominent at smaller angles. The dashed lines in Fig.\ 5a represent a global {\color{red}numerical} fit to the spectra with the parameters (appropriate for 100 K) $\mu_0H_{E}$ = 0.222(2) T, $\mu_0H_{a}$ = 0.147(2) T, and $\mu_0H_{c}$ = 0.625(7) T.

\begin{figure}[htpb]
    \centering
    \includegraphics[width=1.0\textwidth]{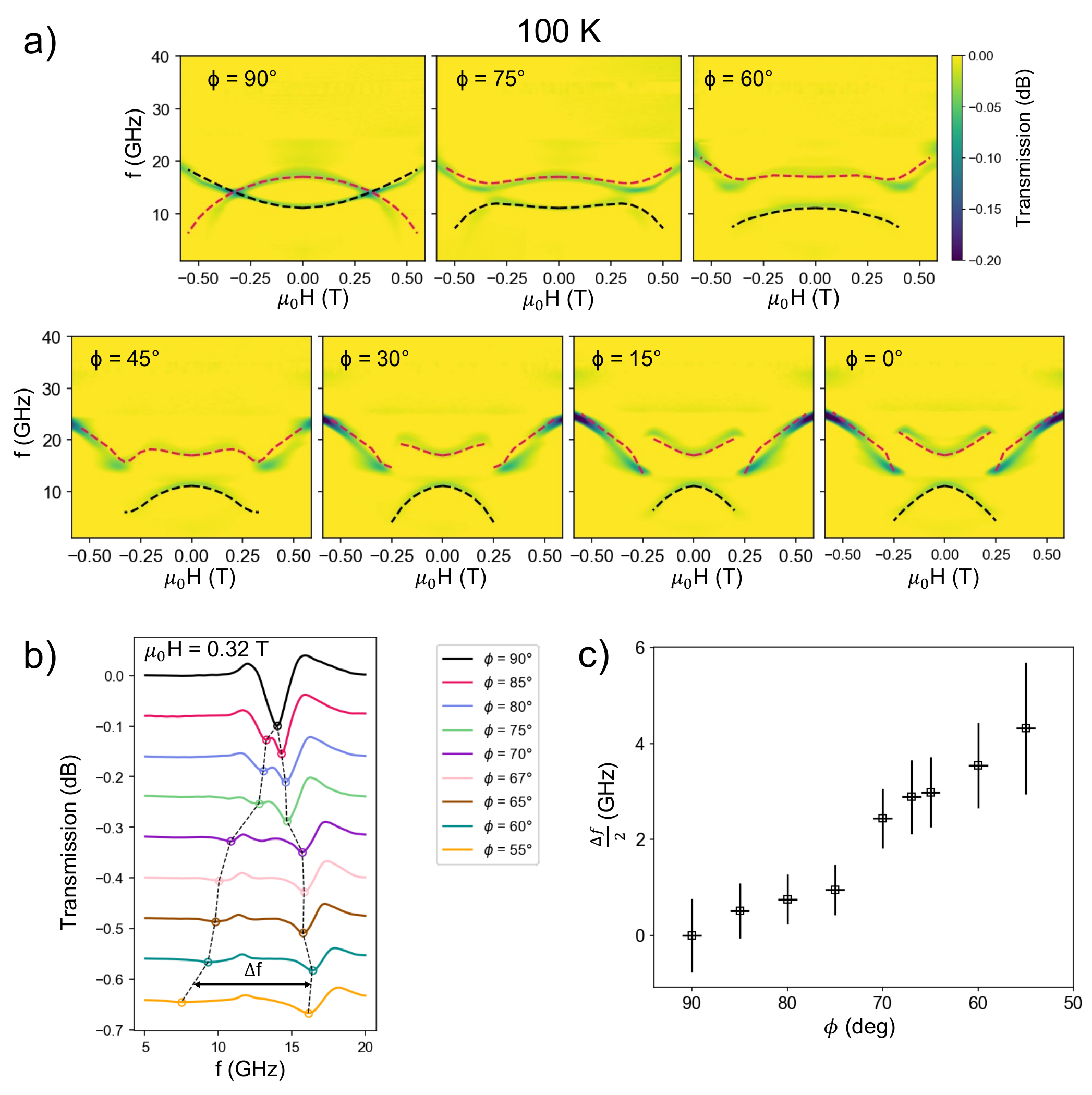}
    \label{Figure5}
    \caption{\textbf{Evolution of a mode crossing at 100 K as a function of 
orienting the in-plane applied magnetic field away from $\boldsymbol{H_{\perp}}$.} a) Evolution of the absorption spectra at 100 K for intermediate angles of in-plane applied magnetic field between $H_{\perp}$ ($\phi$ = 90$^\circ$) and $H_{\parallel}$ ($\phi$ = 0$^\circ$). The dashed lines are determined by fitting the data for $\phi$ = 0$^\circ$ and 90$^\circ$ to eqs 2-6 to obtain the values of the fit parameters $H_E$, $H_a$, and $H_c$ discussed in the text, and then calculating the resonance frequencies numerically for the intermediate angles. b) Line cuts of the absorption spectra near the avoided crossing for different applied magnetic-field angles $\phi$ = 90$^\circ$ to 55$^\circ$, for an applied field magnitude of $\mu_0 H$ = 0.32 T. Plots are offset vertically for clarity. c) Plot of coupling strength $\frac{\Delta f}{2}$ calculated as half the separation between absorption resonances, as a function of $\phi$.}
\end{figure}

\vspace{0.5cm}
Line scans showing the development of the anti-crossing gap in the spectra are shown in Fig.\ 5b for small angles away from $\phi = 90^\circ$. We quantify in Fig.\ 5c the strength of the mode coupling with the parameter $\frac{\Delta f}{2}$, half the frequency separation of the coupled resonance mode peaks. The coupling parameter increases monotonically as the external field is rotated from 90$^\circ$ to 55$^\circ$. For $\phi$ = 55$^\circ$, the coupling strength is $\frac{\Delta f}{2}$ = 4(1) GHz, while the full-width half maxima of the upper and lower modes are $\frac{K_u}{2\pi} \approx 1.14$ GHz and $\frac{K_l}{2\pi} \approx 1.6$ GHz. This satisfies the condition $\frac{\Delta f}{2}$ $>$ $K_u$ and $\frac{\Delta f}{2}$ $>$ $K_l$, indicating that the system is in the regime of strong magnon-magnon coupling, as has previously been observed in other spin-wave systems \cite{Chen2018, Macneill2019, Liensberger2019}.

\vspace{0.5cm}
In summary, we report measurements of  GHz-frequency antiferromagnetic resonance modes in the van der Waals antiferromagnet CrSBr that are anisotropic with regard to the angle of applied magnetic field relative to the crystal axes.  The modes are well described by two coupled Landau-Lifshitz equations for modeling the spin sublattices, when one accounts for interlayer exchange and triaxial magnetic anisotropy present in CrSBr. The interlayer exchange and the anisotropy field strengths were measured by fitting the resonances for a series of temperatures from 5 to 128 K. At 5 K we determine an interlayer exchange field $H_E$ = 0.395(2) T, and aniosotopy fields $H_a$ = 0.383(7) T and $H_c$ =  1.30(2) T.  All three parameters weaken with increasing temperature. As the angle of an applied magnetic field is changed within the $a$-$b$ crystal plane, we observe a continuous evolution from unhybridized acoustic and optical modes with a simple mode crossing (for $H$ parallel to the crystal $a$ axis) to hybridized superpositions with an avoided crossing that is controllable by magnetic field (as the angle of $H$ is tuned away from $a$ axis). This evolution of strong magnon-magnon coupling can be understood as a consequence of breaking a two-fold rotational symmetry that is present when the applied field is aligned along the crystal $a$ axis. Our characterization of antiferromagnetic resonances in an easily-accessible frequency range, and the understanding of how the resonances can be tuned between uncoupled and strongly-coupled by adjusting magnetic field, sets the stage for future experiments regarding  manipulation of the modes and the development of capabilities like antiferromagnetic spin-torque nano-oscillators.

\section{Crystal growth and characterization}
CrSBr single crystals were synthesized using a modified chemical vapor transport approach
adapted from the original report by Beck \cite{Beck1990}. Disulfur dibromide and chromium metal were added together in a 7:13 molar ratio to a fused silica tube approximately 35 cm in length, which was sealed under vacuum and placed in a three-zone tube furnace. The tube was heated in a temperature gradient (1223 to 1123 K) for 120 hours. CrSBr grows as black, shiny flat needles. The long axis of bulk needle crystals of CrSBr has been correlated to the a crystal axis by XRD experiments.

\section{Supporting Information}
Experimental methods, calculations of antiferromagnetic resonances from the Landau-Lifshitz equation, full temperature series, symmetry analyses, data from a second crystal.

\section{Acknowledgements}
We acknowledge inspiring discussions with Kin Fai Mak, Jie Shan, Joseph Mittelstaedt, Kihong Lee, and Caitlin Carnahan. The research at Cornell was supported by the AFOSR/MURI project 2DMagic (FA9550-19-1-0390) and the US National Science Foundation (DMR-2104268). T.M.J.C. was supported by the Singapore Agency for Science, Technology, and Research, and Y.K.L. acknowledges the Cornell Presidential Postdoctoral Fellowship. The work utilized the shared facilities of the Cornell Center for Materials Research (supported by the NSF via grant DMR-1719875) and the Cornell NanoScale Facility, a member of the National Nanotechnology Coordinated Infrastructure (supported by the NSF via grant NNCI-2025233), and it benefited from instrumentation support by the Kavli Institute at Cornell. Synthesis of the CrSBr crystals was supported as part of Programmable Quantum Materials, an Energy Frontier Research Center funded by the U.S. Department of Energy (DOE), Office of Science, Basic Energy Sciences (BES), under award DE-SC0019443.

\section{Author Contributions}
T.M.J.C and Y.K.L devised the experiment and performed the measurements. T.M.J.C performed the data analysis with assistance from Y.K.L.. S.K. assisted with experimental setup and L-L modeling. A.H.D. synthesized the crystals under the supervision of X.R.. D.C.R. provided oversight and advice. T.M.J.C, Y.K.L, and D.C.R. wrote the manuscript. All authors discussed the results and the content of the manuscript.

\section{Competing Interests}
The authors declare no competing interests.

\section{Present Addresses}
\noindent\textsuperscript{\#}S.K.: IBM T. J. Watson Research Center, Yorktown Heights, NY, 10598, USA

\bibliography{references.bib}



\section*{Supplementary material}

# Supplement to "Anisotropic Gigahertz Antiferromagnetic Resonances of the Easy-Axis van der Waals Antiferromagnet CrSBr"

Thow Min Jerald Cham,[1,*] Saba Karimeddiny,[1] Avalon H. Dismukes,[2] Xavier Roy,[2] Daniel C. Ralph,[1,3,†] and Yunqiu Kelly Luo[1,3,4,*]

[1]*Cornell University, Ithaca, NY 14850, USA*
[2]*Department of Chemistry, Columbia University, New York, NY 10027, USA*
[3]*Kavli Institute at Cornell, Ithaca, NY 14853, USA*
[4]*Department of Physics and Astronomy,*
*University of Southern California, Los Angeles, CA 90089, USA*

(Dated: August 5, 2022)

[*] These authors contributed equally
[†] Correspondence email address: yl664@cornell.edu, kelly.y.luo@usc.edu, dcr14@cornell.edu

**CONTENTS**

# I. ACQUISITION OF THE MICROWAVE ABSORPTION SPECTRA

We couple the CrSBr crystal to a ground-signal-ground coplanar waveguide (CPW) and perform RF (1 - 40 GHz) transmission measurements. For the data analyzed in the main text, we aligned the long edge of the crystal (a-axis) perpendicular to the middle signal line and glued it in place using rubber cement, as shown in Fig. S1(a). We also took special care to align the vdW plane (CrSBr a-b crystal plane) parallel with the waveguide plane. We connected the ends of the CPW to ports 1 and 2 of a Keysight 8722ES 40 GHz S-parameter vector network analyzer (VNA) using coaxial cables. The waveguide was then lowered into a Janis He vapour flow cryostat and secured onto a brass sample stage containing a sample thermometer. The temperature was set with a Lakeshore 331 temperature PID controller. We used a GMW model 5403 electromagnet, mounted on an stepper-motor-controlled in-plane rotation stage, to sweep the external field $H_{ext}$ parallel ($H_\parallel$) or perpendicular ($H_\perp$) with respect to the Néel axis (crystal b-axis), or at angles in between.

We also performed transmission measurements on another crystal, for which we aligned the long a-axis parallel to the middle signal line of the CPW, rather than perpendicular. We report results from that sample in Section VI of this supplementary information.

Before acquiring spectra at each temperature, we wait for the temperature to stabilize and then calibrate the VNA using open, short, and 50 Ω standards to account for temperature-dependent transmission factors. We measure the magnitude of the $S_{21}$ transmission vector, subtract a field-averaged background for each frequency, and perform Fourier Transform smoothing and interpolation to remove any streak-like artifacts in the spectra. The transmission spectrum before and after the subtraction and smoothing process at 5 K are shown in Fig. S1(b) and (c).

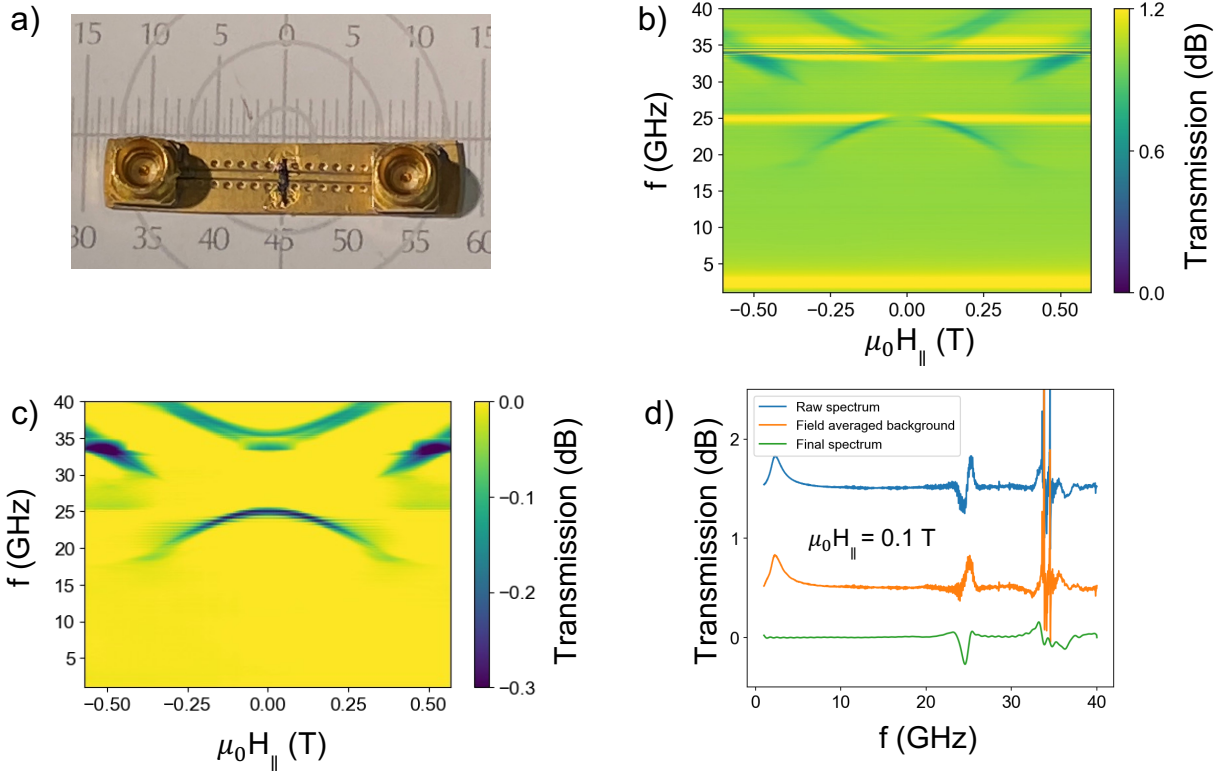

FIG. S1: Details of the microwave absorption measurements. a) CrSBr crystal with the $a$ axis aligned perpendicular to the signal line of a ground-signal-ground coplanar waveguide. Absorption spectra b) before and c) after the background subtraction and smoothing procedure. d) Comparison of spectra linecuts at $\mu_0 H = 0.1$ T with the field-averaged background. Plots are offset vertically for clarity.

## II. CALCULATION OF ANTIFERROMAGNETIC RESONANCES FOR CrSBr SAMPLES ACCOUNTING FOR TRIAXIAL ANISOTROPY

To model the antiferromagnetic resonance frequencies, we consider the coupled dynamics of the two spin sublattices, 1 and 2. In the approximation of small-angle precession, we can write the magnetization orientation of each sublattice as a unit vector consisting of a

time-invariant equilibrium position plus a small time-varying part:

$$\hat{\mathbf{m}}_{1(2)} = \hat{\mathbf{m}}_{1(2)}^{eq} + \delta\hat{\mathbf{m}}_{1(2)}e^{i\omega t}. \quad (1)$$

In our analysis, we define the $\hat{c}$ direction to be perpendicular to the van der Waal layers and along CrSBr hard axis ($c$-axis); $\hat{a}$ to be along the CrSBr intermediate axis ($a$-axis); and $\hat{b}$ to be along the CrSBr easy axis ($b$-axis). We assume that the exchange interaction felt by one sublattice imposed by the other sublattice can be written as $-H_E\hat{\mathbf{m}}_{2(1)}$. Therefore, the general expression for the total effective magnetic field (including the external field, exchange, and first-order anisotropy terms) is

$$\boldsymbol{H}_{1(2)} = H_b^{ext}\hat{b} + (H_a^{ext} - H_a(m_{1(2)a}^{eq} + \delta m_{1(2)a})e^{i\omega t})\hat{a} \\ + (H_c^{ext} - H_c(m_{1(2)c}^{eq} + \delta m_{1(2)c})e^{i\omega t})\hat{c} - H_E\hat{\mathbf{m}}_{2(1)}, \quad (2)$$

where $H_a$ and $H_c$ are anisotropy parameters along the CrSBr intermediate and hard axes [1, 2].

The general Landau-Lifshitz (L-L) equation without damping is written as:

$$\frac{d\hat{\mathbf{m}}_{1(2)}}{dt} = -\mu_0\gamma\hat{\mathbf{m}}_{1(2)} \times \boldsymbol{H}_{1(2)}. \quad (3)$$

### A. External field applied along the CrSBr intermediate axis ($H_\perp$)

For the case when the extermal field is applied along the $a$ axis, it induces a spin-flop configuration such that the equilibrium sublattice magnetizations $\hat{\mathbf{m}}_{1(2)}$ stay in the $b - a$ plane and tilt symmetrically toward $\hat{a}$, as illustrated in Fig. S2 below. Therefore, $\hat{\mathbf{m}}_{1(2)}^{eq}$ can be written as $(m_{1(2)b}^{eq}, m_{1(2)a}^{eq}, m_{1(2)c}^{eq}) = (\pm\cos\chi, \sin\chi, 0)$.

Substituting $\mu_0 H_b^{ext} = \mu_0 H_c^{ext} = 0$ T and $\mu_0 H_a^{ext} = \mu_0 H_\perp$ into Eq. (2), the total effective magnetic field can be written as

$$\boldsymbol{H}_{1(2)} = (H_\perp - H_a(m_{1(2)a}^{eq} + \delta m_{1(2)a})e^{i\omega t})\hat{a} - H_c(m_{1(2)c}^{eq} + \delta m_{1(2)c}e^{i\omega t})\hat{c} - H_E\hat{\mathbf{m}}_{2(1)}. \quad (4)$$

Substituting in the equilibrium spin-flop configuration $\hat{\mathbf{m}}_{1(2)}^{eq} = (\pm\cos\chi, \sin\chi, 0)$, Eq. (4)

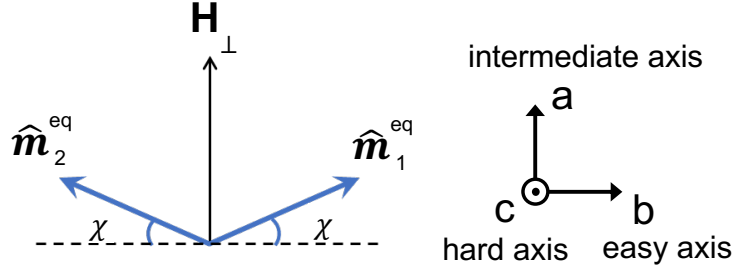

FIG. S2: Configuration of antiferromagnetic sub-lattices with $H_{ext}$ along the intermediate axis. This configuration is symmetric upon two-fold rotation about the $\hat{a}$-axis and interchange of the spin sub-lattices.

becomes

$$\boldsymbol{H}_{1(2)} = (-H_E(\delta m_{2(1)b}e^{i\omega t} \mp \cos\chi))\hat{b} + (H_\perp - H_a(\delta m_{1(2)a}e^{i\omega t} + \sin\chi) - H_E(\delta m_{2(1)a}e^{i\omega t} + \sin\chi))\hat{a}$$
$$+ (-H_c\delta m_{1(2)c}e^{i\omega t} - H_E\delta m_{2(1)c}e^{i\omega t})\hat{c}. \tag{5}$$

Therefore, writing the L-L equations (Eq. (3)) in matrix form and ignoring terms higher than first-order in the precession amplitude:

$$\begin{vmatrix} -iw & 0 & \gamma[H_\perp + \sin\chi(H_c - H_a - H_E)] & 0 & 0 & \gamma\sin\chi H_E \\ 0 & -iw & -\gamma\cos\chi(H_c + H_E) & 0 & 0 & -\gamma\cos\chi H_E \\ \gamma[-H_\perp + \sin\chi(H_a + H_E)] & \gamma\cos\chi(H_a + H_E) & -iw & -\gamma\sin\chi H_E & \gamma\cos\chi H_E & 0 \\ 0 & 0 & \gamma\sin\chi H_E & -iw & 0 & \gamma[H_\perp + \sin\chi(-H_a + H_c - H_E)] \\ 0 & 0 & \gamma\cos\chi H_E & 0 & -iw & \gamma\cos\chi(H_E + H_c) \\ -\gamma\sin\chi H_E & -\gamma\cos\chi H_E & 0 & \gamma[-H_\perp + \sin\chi(H_E + H_a)] & \gamma\cos\chi(-H_a - H_E) & -iw \end{vmatrix} = 0 \tag{6}$$

To obtain the equilibrium position, since the moments will remain in the $b-a$ plane, we can ignore the $c$-axis anisotropy and write the free energy (per unit sub-lattice magnetization) as

$$F = -\hat{\mathbf{m}}_1 \cdot \boldsymbol{H}^{ext} - \hat{\mathbf{m}}_2 \cdot \boldsymbol{H}^{ext} + H_E\hat{\mathbf{m}}_1 \cdot \hat{\mathbf{m}}_2 + \frac{1}{2}H_a(\hat{\mathbf{m}}_1 \cdot \hat{a})^2 + \frac{1}{2}H_c(\hat{\mathbf{m}}_2 \cdot \hat{c})^2, \tag{7}$$

and then minimize with respect to $\chi$. For an applied magnetic field perpendicular to the easy axis, this gives $\sin\chi = H_\perp/(2H_E + H_a)$. We substitute this expression into the $6 \times 6$

eigenvalue equation (6) and solve it to get 6 eigenvalues. Only two of these are non-trivial and positive, corresponding to the resonance frequencies $\omega_{1(2),\perp}$:

$$\omega_{1,\perp} = \mu_0 \gamma \sqrt{\frac{(H_\perp^2(2H_E - H_a) + H_a(H_a + 2H_E)^2)(2H_E + H_c)}{(H_a + 2H_E)^2}} \tag{8}$$

$$\omega_{2,\perp} = \mu_0 \gamma \sqrt{\frac{((H_a + 2H_E)^2 - H_\perp^2)H_c}{H_a + 2H_E}}. \tag{9}$$

These resonance modes are generalizations of the acoustic and optical modes previously reported for CrCl$_3$ [3]. By considering the two-fold rotational symmetry of the system about the $a$ axis, we find that the $\omega_{1,\perp}$ mode is odd with respect to this symmetry and so it will be excited by RF fields that are odd under a 2-fold rotation about $\hat{a}$. The $\omega_{2,\perp}$ mode is even and will be excited by RF fields that have an even component.

At large applied magnetic fields for which the two sublattices are driven completely parallel (For $H_\perp \geq 2H_E + H_a$), $\hat{\mathbf{m}}_{1(2)}^{eq} = (0, 1, 0)$, $\chi \to \pi/2$, Eqs. (8) and (9) are no longer applicable, and we get the uniform ferromagnetic resonance mode

$$\omega_{1,\perp,FM} = \mu_0 \gamma \sqrt{(H_\perp - H_a)(H_\perp - H_a + H_c)} \tag{10}$$

and a lower-frequency mode

$$\omega_{2,\perp,FM} = \mu_0 \gamma \sqrt{(H_\perp - H_a - 2H_E)(H_\perp - H_a + H_c - 2H_E)} \tag{11}$$

In the lower-frequency mode, $\delta\hat{\mathbf{m}}_1 = -\delta\hat{\mathbf{m}}_2$ so there is no net time-dependent magnetization, and for this reason we suggest that this mode will not couple to an external RF magnetic field or produce any signal in a CPW measurement.

### B. External field applied along the CrSBr easy axis ($H_\parallel$)

When an external magnetic field is applied along the CrSBr easy axis ($b$-axis), there are three possible equilibrium configurations, depending on the relative values of the $H_\parallel$ (the applied field), $H_E$, and $H_a$. The two spin sublattices may remain antiparallel and aligned along the easy axis, in which case according to Eq. (7) the energy (per unit sub-lattice

magnetization) of the state is $F_{AP} = -H_E$. The Néel vector may rotate perpendicular to the applied field with the spin sublattices forming a spin-flop state with the canting angle $\sin\chi = H_\parallel/(2H_E - H_a)$, resulting in the energy $F_{\text{flop}} = -H_\parallel^2/(2H_E - H_a) - H_E + H_a$. Alternatively, the two spin sublattices can be driven parallel to the applied magnetic field, with the energy $F_P = H_E - 2H_\parallel$. The equilibrium configuration can be determined based on which state minimizes the energy. The field values corresponding to the various transitions are as follows: the spin-flop state becomes lower in energy than the antiparallel state when $H_\parallel > \sqrt{(2H_E - H_a)H_a}$, the parallel state becomes lower in energy than the antiparallel state when $H_\parallel > H_E$, and the canting angle of the spin-flop state goes to $\pi/2$ to reach the parallel state when $H_\parallel > 2H_E - H_a$. This results in two possible scenarios for the evolution of the configurations of the antiferromagnet, as pictured in Fig.S3.

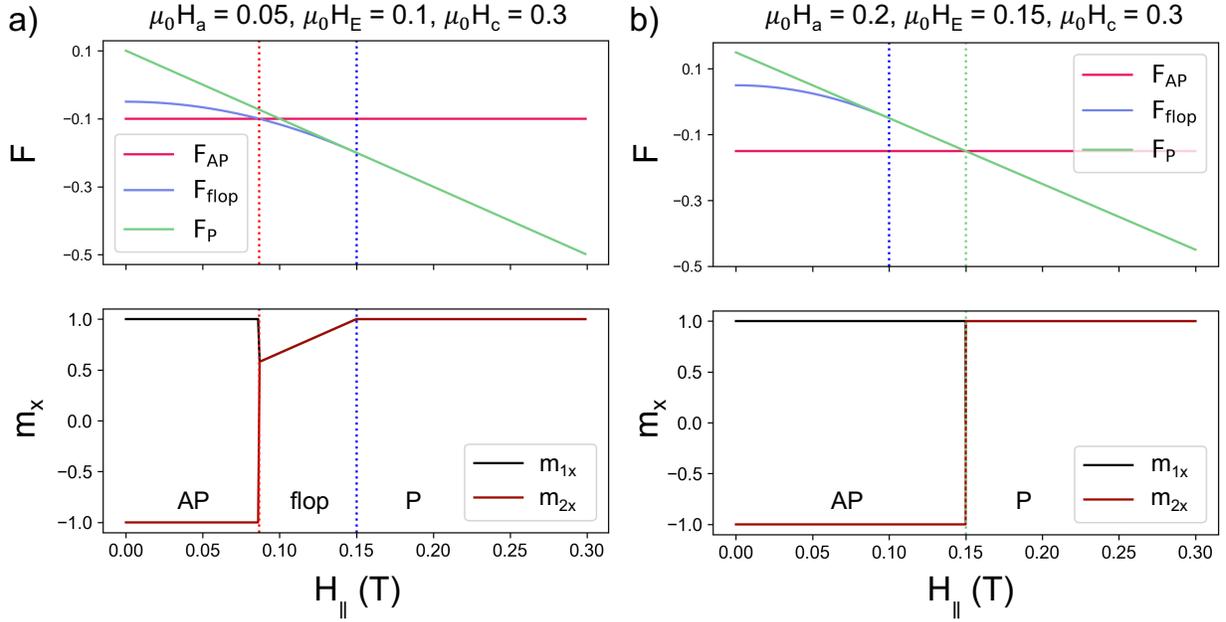

FIG. S3: Field-induced magnetic phase transitions for scenario (a) weak anisotropy: spin-flop transition at intermediate applied field $H_\parallel > \sqrt{(2H_E - H_a)H_a}$ (dashed red line) followed by a gradual transition into a parallel state when $H_\parallel > 2H_E - H_a$ (dashed blue line); and for scenario (b) strong anisotropy: spin-flip transition at $H_\parallel > H_e$ (dashed green line).

8

I. If the intermediate-axis anisotropy is relatively weak ($H_a < H_E$), the antiferromagnet is initially in the antiparallel state at low values of applied field, and as a function of increasing field it first makes a sudden, discontinuous transition to the spin-flop state at $H_\parallel = \sqrt{(2H_E - H_a)H_a}$, and then the canting angle grows continuously until the fully-parallel state is reached for fields $H_\parallel > 2H_E - H_a$ or greater.

II. If $H_a > H_E$, as a function of increasing $H_\parallel$ there is no value of field for which the spin-flop state is stable. The antiferromagnet is initially in the antiparallel state for low values of applied field, but then makes a discontinuous transition to the fully-parallel state for $H_\parallel > H_E$.

We will calculate the resonance frequencies expected for all three states (antiparallel, spin-flop, and fully-parallel), and compare to our measurements. For CrSBr we will find that $H_a$ is less than $H_E$, but the values at most temperatures are rather close, so there is only a narrow range of field for which the spin-flop state is stabilized. Therefore the resonances that are most prominent in the measured spectra correspond to the antiparallel and fully-parallel states.

The same type of L-L analysis discussed above can be used to obtain analytic solutions for the resonance modes corresponding to each of the possible equilibrium configurations. Since here we are considering only applied magnetic fields parallel to $\hat{b}$, the easy-axis, $\mu_0 H_a^{ext} = \mu_0 H_c^{ext} = 0$ T and $\mu_0 H_b^{ext} = \mu_0 H_\parallel$. The total effective magnetic field given by Eq. (2) becomes

$$\boldsymbol{H}_{1(2)} = H_\parallel \hat{b} - H_a(m_{1(2)a}^{eq} + \delta m_{1(2)a})e^{i\omega t}\hat{a} - H_c(m_{1(2)c}^{eq} + \delta m_{1(2)c})e^{i\omega t}\hat{c} - H_E \hat{\mathbf{m}}_{2(1)}. \quad (12)$$

*1. Zero to small applied field, before spin-flop or spin-flip transition*

Here the equilibrium orientations of both spin sublattices remain aligned with the $\hat{b}$ easy axis, so $\hat{\mathbf{m}}_{1(2)}^{eq} = (\pm 1, 0, 0)$. With this, Eq. (12) reduces to

$$\boldsymbol{H}_{1(2)} = (H_\parallel - H_E)\hat{b} - H_a \delta m_{1(2)a}e^{i\omega t}\hat{a} - H_c \delta m_{1(2)c}e^{i\omega t}\hat{c}, \quad (13)$$

9

Expanding the L-L equations (Eq. 3) and taking terms to the first order in the precession amplitude gives the $6 \times 6$ eigenvalue equation

$$\begin{vmatrix} -iw & 0 & 0 & 0 & 0 & 0 \\ 0 & -iw & -\gamma(H_\| + H_c + H_E) & 0 & 0 & -H_E\gamma \\ 0 & \gamma[H_\| + H_a + H_E)] & -iw & 0 & H_E\gamma & 0 \\ 0 & 0 & 0 & -iw & 0 & 0 \\ 0 & 0 & H_E\gamma & 0 & -iw & \gamma(-H_\| + H_c + H_E) \\ 0 & -H_E\gamma & 0 & 0 & \gamma(H_\| - H_a - H_E) & -iw \end{vmatrix} = 0. \tag{14}$$

Solving this, we again obtain two positive non-trivial resonance frequencies:

$$\omega_{1,\|} = \mu_0\gamma\sqrt{H_\|^2 + H_a(H_E + H_c) + H_E H_c - \sqrt{H_\|^2(H_a + H_c)(H_a + 4H_E + H_c) + H_E^2(H_a - H_c)^2}} \tag{15}$$

$$\omega_{2,\|} = \mu_0\gamma\sqrt{H_\|^2 + H_a(H_E + H_c) + H_E H_c + \sqrt{H_\|^2(H_a + H_c)(H_a + 4H_E + H_c) + H_E^2(H_a - H_c)^2}}. \tag{16}$$

These modes are linear superpositions of the left-handed (LH) and right-handed (RH) modes depicted in Fig. 2 in the main text [2, 4] that would exist for an antiferromagnet with purely uniaxial anisotropy. The broken rotational symmetry about the field axis in a material with triaxial anisotropy ($H_c \neq H_a$) causes the LH and RH modes to be hybridized.

2. *Spin-flop configuration*

Using the spin-flop configuration $\hat{\mathbf{m}}_{1(2)}^{eq} = (\sin\chi, \pm\cos\chi, 0)$, Eq. (12) can be written as

$$\boldsymbol{H}_{1(2)} = [H_\| - H_E(\delta m_{2(1)b}e^{i\omega t} + \sin\chi)]\hat{b} + [-H_a(\delta m_{1(2)a}e^{i\omega t} \pm \cos\chi) - H_E(\delta m_{2(1)a}e^{i\omega t} \mp \cos\chi)]\hat{a}$$
$$+ [-H_c\delta m_{1(2)c}e^{i\omega t} - H_E\delta m_{2(1)c}e^{i\omega t}]\hat{c} \tag{17}$$

Substituting these expressions into the L-L equations (Eq. (3)), removing the equilibrium terms, and ignoring the higher-order terms in the precession amplitudes gives:

$$\begin{vmatrix} iw & 0 & \gamma\cos\chi(H_c - H_a + H_E) & 0 & 0 & \gamma\cos\chi H_E \\ 0 & iw & -\gamma[H_\| + \sin\chi(H_c - H_E)] & 0 & 0 & -\gamma\sin\chi H_E \\ \gamma[\cos\chi(H_a - H_E)] & \gamma[H_\| + \sin\chi(H_a - H_E)] & iw & -\gamma\cos\chi H_E & \gamma\sin\chi H_E & 0 \\ 0 & 0 & -\gamma\cos\chi H_E & iw & 0 & \gamma[\cos\chi(H_a - H_c - H_E)] \\ 0 & 0 & -\gamma\sin\chi H_E & 0 & iw & \gamma[-H_\| + \sin\chi(H_E - H_c)] \\ \gamma\cos\chi H_E & \gamma\sin\chi H_E & 0 & \gamma\cos\chi(H_E - H_a) & \gamma[H_\| + \sin\chi(H_a - H_E)] & iw \end{vmatrix} = 0.$$

(18)

Minimizing free energy (Eq. (7)) with respect to $\chi$ gives $\sin\chi = H_\|/(2H_E - H_a)$, and solving for the eigenvalues yields

$$\omega_{1,\|,\text{flop}} = \mu_0\gamma\sqrt{\frac{(H_\|^2(H_a + 2H_E) - H_a(H_a - 2H_E)^2)(2H_E + H_c - H_a)}{(H_a - 2H_E)^2}} \quad (19)$$

$$\omega_{2,\|,\text{flop}} = \mu_0\gamma\sqrt{\frac{((2H_E - H_a)^2 - H_\|^2)(H_c - H_a)}{2H_E - H_a}}. \quad (20)$$

where $\omega_{1,\|,\text{flop}}$ corresponds to an acoustic mode and $\omega_{2,\|,\text{flop}}$ corresponds to an optical mode.

3. *Parallel orientation of the spin sub-lattices*

For applied fields $H_\|$ sufficiently strong that the two spin sublattices become parallel to each other, $\hat{\mathbf{m}}_{1(2)}^{eq} = (1, 0, 0)$, $\chi \to \pi/2$. Analogous to the results in Eqs. (10) and (11), we get the usual ferromagnetic mode

$$\omega_{1,\|,FM} = \mu_0\gamma\sqrt{(H_\| + H_a)(H_\| + H_c)} \quad (21)$$

and a lower frequency mode

$$\omega_{2,\|,FM} = \mu_0\gamma\sqrt{(H_\| + H_a - 2H_E)(H_\| + H_c - 2H_E)} \quad (22)$$

## C. Spectra fitting

To determine the exchange and anisotropy parameters, we fit the measured spectra to the appropriate piece-wise continuous functions. For magnetic field ($H_\perp$) applied parallel to the $a$ axis (the intermediate magnetic axis), the fitting function is

$$f_\perp(H_\perp) = \begin{cases} (8), (9) & \text{for } H_\perp < 2H_E + H_a \\ (10) & \text{for } H_\perp > 2H_E + H_a \end{cases} \quad (23)$$

For a magnetic field ($H_\parallel$) applied parallel to the $b$ axis (the easy magnetic axis), if $H_a < H_E$:

$$f_\parallel(H_\parallel) = \begin{cases} (15), (16) & \text{for } H_\parallel < \sqrt{(2H_E - H_a)H_a} \\ (19) & \text{for } \sqrt{(2H_E - H_a)H_a} < H_\parallel < 2H_E - H_a \\ (21) & \text{for } H_\parallel > 2H_E - H_a \end{cases} \quad (24)$$

and for the alternative case if $H_a > H_E$:

$$f_\parallel(H_\parallel) = \begin{cases} (15), (16) & \text{for } H_\parallel < H_E \\ (21) & \text{for } H_\parallel > H_E. \end{cases} \quad (25)$$

As shown in Fig. S4, the fit parameters obtained from the piecewise fitting procedure give $H_a < H_E$ for all temperatures, such that in our model the spin-flop configuration should be stabilized for the field range $\sqrt{H_a(2H_E - H_a)} < H < 2H_E - H_a$. At the base temperature of 5 K, $H_a$ is only slightly less than $H_E$ so that there should be only a very small field window exhibiting the spin-flop configuration, and the spectra is almost equivalent to that of a spin-flip transition. With increasing temperature, the anisotropy strength decreases faster than the exchange strength, resulting in a more prominent spin-flop regime.

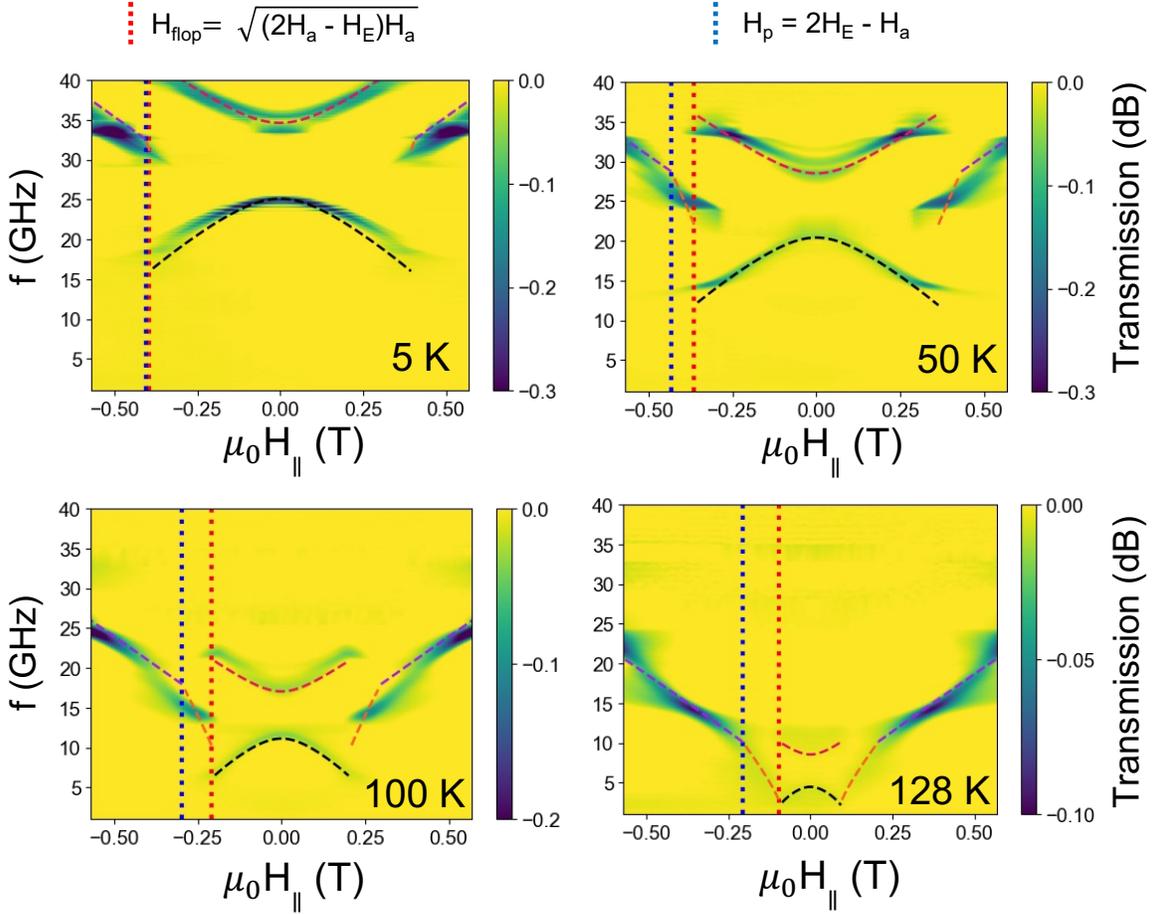

FIG. S4: Spectra from Sample 1 of the main text, with dashed lines representing the fit functions discussed in the main text and supplementary information, Section II. The vertical red line represents the predicted spin-flop transition field, and the blue line represents the field where the canting angle in the spin-flop state should realize $\chi = \pi/2$ to achieve fully-parallel spin alignment.

## III. EFFECT OF TRIAXIAL ANISOTROPY ON ANTIFERROMAGNETIC RESONANT MODES

Here we analyze in more detail the effects of triaxial magnetic anisotropy on antiferromagnetic resonance modes, comparing to both the easy-plane anisotropy case and the uniaxial easy-axis case.

13

For an easy-plane antiferromagnet, with external field applied within the easy plane, the normal modes in the spin-flop configuration can be expressed as [3]

$$\omega_{1,\perp} = \mu_0 \gamma H_\perp \sqrt{1 + \frac{H_c}{2H_E}} \tag{26}$$

$$\omega_{2,\perp} = \mu_0 \gamma \sqrt{\frac{(4H_E^2 - H_\perp^2)H_c}{2H_E}}, \tag{27}$$

where $\omega_{1,\perp}$ corresponds to the acoustic mode and $\omega_{2,\perp}$ the optical mode, e.g., as measured previously in CrCl$_3$ [3]. These expressions are equivalent to the $H_\perp$ modes given by Eqs. (8) and (9) when $\mu_0 H_a = 0$ T. We plot these frequencies as solid lines in Fig. S5(a) using parameters appriate for CrSBr except that we set $\mu_0 H_a = 0$ T. With the addition of an in-plane intermediate anisotropy energy along the a-axis ($\mu_0 H_a \neq 0$ T), both acoustic and optical modes increase in frequency (dashed lines in Fig. S5(a)). The intermediate anisotropy axis also causes the dispersion of the acoustic mode to deviate from a linear relation with field and gaps this mode for low applied magnetic fields.

For an antiferromagnet with uniaxial anisotropy and external field along the preferred axis, the normal modes are [2]

$$\omega_{1,\parallel} = \mu_0 \gamma \sqrt{H_\parallel^2 + 2H_c H_E + H_c^2 - 2H_\parallel \sqrt{H_c(2H_E + H_c)}} \tag{28}$$

$$\omega_{2,\parallel} = \mu_0 \gamma \sqrt{H_\parallel^2 + 2H_c H_E + H_c^2 + 2H_\parallel \sqrt{H_c(2H_E + H_c)}} \tag{29}$$

where $\omega_{1,\parallel}$ and $\omega_{2,\parallel}$ correspond to the left-handed (LH) and right-handed (RH) modes, e.g. as measured previously in Cr$_2$O$_3$ [4]. These frequencies are equivalent to the $H_\parallel$ modes given by Eqs. (15) and (16), when $H_a = H_c$. The two modes described by Eqs. (28), (29) are degenerate at $\mu_0 H_\parallel = 0$ T and disperse linearly at low field (solid lines in Fig. S5(b). Similar to gadolinium iron garnet [5], the presence of triaxial anisotropy for CrSBr ($H_a \neq H_c$), breaks rotational symmetry about the easy axis and couples the RH and LH modes to lift the degeneracy at zero applied field (dashed lines fitted to the measured spectrum in Fig.S5(b)).

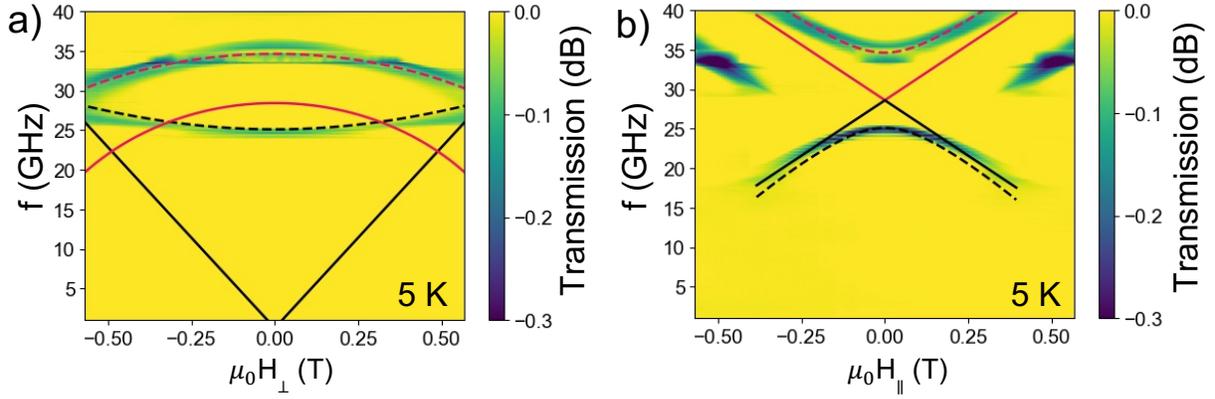

FIG. S5: a) Comparison of resonance frequencies assuming easy-plane anistropy ($\mu_0 H_a = 0$ T, solid lines) vs. the measured modes in CrSBr for magnetic field applied in-plane perpendicular to the easy axis at 5 K, along with fits (dashed lines) assuming triaxial anisotropy. b) Comparison of resonance frequencies assuming uniaxial easy-axis anisotropy ($\mu_0 H_a = \mu_0 H_c = 0.7$ T, solid lines) vs. the measured modes in CrSBr for applied magnetic field parallel to the easy axis at 5 K, along with fits (dashed lines) assuming triaxial anisotropy. The fit parameters are $\mu_0 H_a = 0.383(7)$ T, $\mu_0 H_E = 0.395(2)$ T, and $\mu_0 H_c = 1.30(2)$ T.

## IV. RESONANCE MODES FOR WEAK MAGNETIC FIELDS AT ARBITRARY IN-PLANE ANGLES RELATIVE TO THE ANISOTROPY AXES

Here we consider the evolution of the modes when the external field is applied at an arbitrary angle $\phi$ between the $\hat{b}$ and $\hat{a}$ easy and intermediate axes. In this case, $\boldsymbol{H}_{ext} = H_0(\cos\phi, \sin\phi, 0)$ and $\hat{\mathbf{m}}^{eq}_{1(2)} = (\pm\cos\chi_{1(2)}, \sin\chi_{1(2)}, 0)$.

The effective magnetic field acting on each sub-lattice is

$$\boldsymbol{H}_{1(2)} = [H_0\cos\phi - H_E(\delta m_{2(1)b}e^{i\omega t} \mp \cos\chi_{2(1)})]\hat{\boldsymbol{b}}$$
$$+ [H_0\sin\phi - H_a(\delta m_{1(2)a}e^{i\omega t} + \sin\chi_{1(2)}) - H_E(\delta m_{2(1)a}e^{i\omega t} + \sin\chi_{2(1)})]\hat{\boldsymbol{a}} \quad (30)$$
$$+ [-H_c\delta m_{1(2)c}e^{i\omega t} - H_E\delta m_{2(1)c}e^{i\omega t}]\hat{\boldsymbol{c}}.$$

15

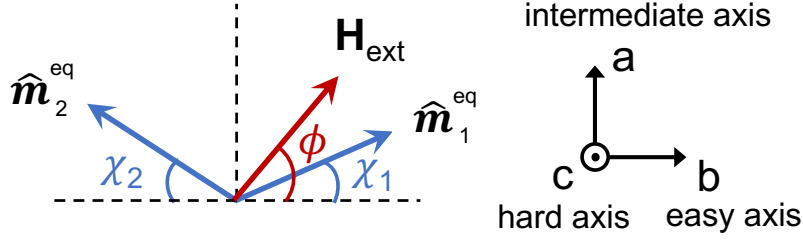

FIG. S6: Equilibrium positions of sub-lattices with external field at an arbitrary angle $\phi$ between the easy and intermediate axes.

At equilibrium, the net force on each sub-lattice is zero [2]

$$H_0 \sin(\phi - \chi_1) - H_E \sin(\chi_1 + \chi_2) - H_a \sin\chi_1 = 0$$
$$H_0 \sin(\phi + \chi_2) - H_E \sin(\chi_1 + \chi_2) - H_a \sin\chi_2 = 0. \quad (31)$$

We consider the small field regime where $H_0 < \sqrt{2 H_a H_E}$, such that $\chi_1$, $\chi_2$ are small. Using the small angle approximation, the equilibrium positions $\chi_{1,2}$ are given by the expression [2]

$$\chi_{1(2)} = \frac{H_0 \sin\phi (H_a \mp H_0 \cos\phi)}{H_a^2 + 2 H_a H_E - H_0^2 \cos^2\phi}. \quad (32)$$

The 6 × 6 matrix encapsulating the coupled L-L equations is

$$\begin{vmatrix} -iw & 0 & M_{13} & 0 & 0 & M_{16} \\ 0 & -iw & M_{23} & 0 & 0 & M_{26} \\ M_{31} & M_{32} & -iw & M_{34} & M_{35} & 0 \\ 0 & 0 & M_{43} & -iw & 0 & M_{46} \\ 0 & 0 & M_{53} & 0 & -iw & M_{56} \\ M_{61} & M_{62} & 0 & M_{64} & M_{65} & -iw \end{vmatrix} = 0. \quad (33)$$

16

with the matrix elements

$$
\begin{aligned}
M_{13} &= -\gamma((H_a - H_c)\sin\chi_1 + H_E \sin\chi_2 - H_0 \sin\phi) \\
M_{16} &= \gamma H_E \sin\chi_1 \\
M_{23} &= -\gamma(H_c \cos\chi_1 + H_E \cos\chi_2 + H_0 \cos\phi) \\
M_{26} &= -\gamma H_E \cos\chi_1 \\
M_{31} &= -\gamma(-H_a \sin\chi_1 - H_E \sin\chi_2 + H_0 \sin\phi) \\
M_{32} &= -\gamma(-H_a \cos\chi_1 - H_E \cos\chi_2 - H_0 \cos\phi) \\
M_{34} &= -\gamma H_E \sin\chi_1 \\
M_{35} &= \gamma H_E \cos\chi_1 \\
M_{43} &= \gamma H_E \sin\chi_2 \\
M_{46} &= -\gamma((H_a - H_c)\sin\chi_2 + H_E \sin\chi_1 - H_0 \sin\phi) \\
M_{53} &= \gamma H_E \cos\chi_2 \\
M_{56} &= \gamma(H_c \cos\chi_2 + H_E \cos\chi_1 - H_0 \cos\phi) \\
M_{61} &= -\gamma H_E \sin\chi_2 \\
M_{62} &= -\gamma H e \cos\chi_2 \\
M_{64} &= \gamma(H_a \sin\chi_2 + H_E \sin\chi_1 - H_0 \sin\phi) \\
M_{65} &= -\gamma(H_a \cos\chi_2 + H_E \cos\chi_1 - H_0 \cos\phi).
\end{aligned}
\tag{34}
$$

At this point we use the small angle approximation for $\chi_{1,2}$ in Eq. (32) and employ matrix diagonalization to obtain an analytical solution for the resonance modes for the small-field regime. Using exchange and anisotropy field strengths obtained from the fits to the experiment at $\phi = 0°$ and $90°$ at 100 K, we plot the evolution of the modes for a range of in-plane angles $\phi = 0°$ to $90°$ in Fig. S7a. We see that the upper and lower modes evolve smoothly from the convex-paired optical and acoustic modes at $\phi = 90°$ to the concave-paired RH and LH modes at $\phi = 0°$. In addition, fields applied at small angles away from $\phi = 90°$ cause the acoustic/optical mode crossing to become gapped. This is analogous to the mode anti-crossing induced by a symmetry breaking out-of-plane field for the resonance modes in easy-plane CrCl$_3$ [3]. In our case, the two-fold rotational symmetry existing for

17

CrSBr with a field applied along the intermediate $\hat{a}$ axis can be broken with an in-plane field component along the easy $\hat{b}$ axis.

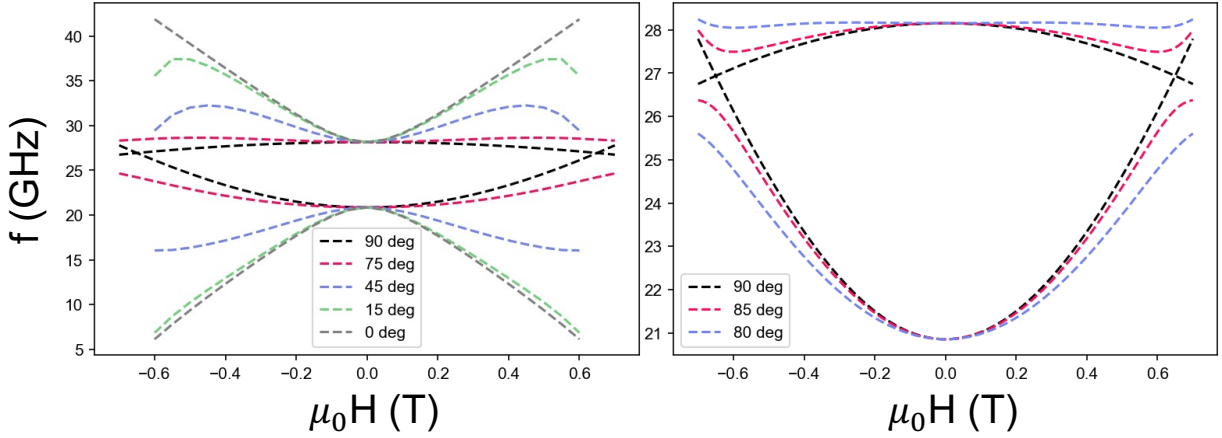

FIG. S7: Resonance modes calculated for external field at an arbitrary angles $\phi$ between the easy and intermediate anisotropy axes, with the parameters (appropriate for 100 K) $\mu_0 H_E = 0.222(5)$ T, $\mu_0 H_a = 0.147(4)$ T, and $\mu_0 H_c = 0.63(1)$ T. The acoustic and optical modes evolve smoothly to the RH and LH modes as the external field is rotated from the intermediate axis ($\phi = 90°$) to the easy axis ($\phi = 0°$). For small angles away from the intermediate axis, we see a gap opening at the mode crossings, associated with the breaking of the two-fold rotation symmetry.

When the external field becomes comparable to the exchange and anisotropy, $H_0 > \sqrt{2 H_a H_E}$, the small angle approximation no longer holds (main text, Fig. 5). Here, we numerically calculate the equilibrium positions $\chi_1$, $\chi_2$ by integrating the L-L equations for a given field direction and field strength. We then use these equilibrium values to evaluate the matrix elements in Eq. (34) and diagonalize the matrix for the resonance mode frequencies.

# V. COMPLETE SERIES OF SPECTRA AT DIFFERENT TEMPERATURES

Here we present the complete series of $H_\perp$ and $H_\parallel$ spectra used to obtain the fit parameters for the exchange and anisotropy fields from 5 - 160 K (Fig. 3 in the main text). Dashed lines represent fits, while the red and blue dotted lines represent the predicted spin-flop and fully-parallel spin alignment transition fields, respectively.

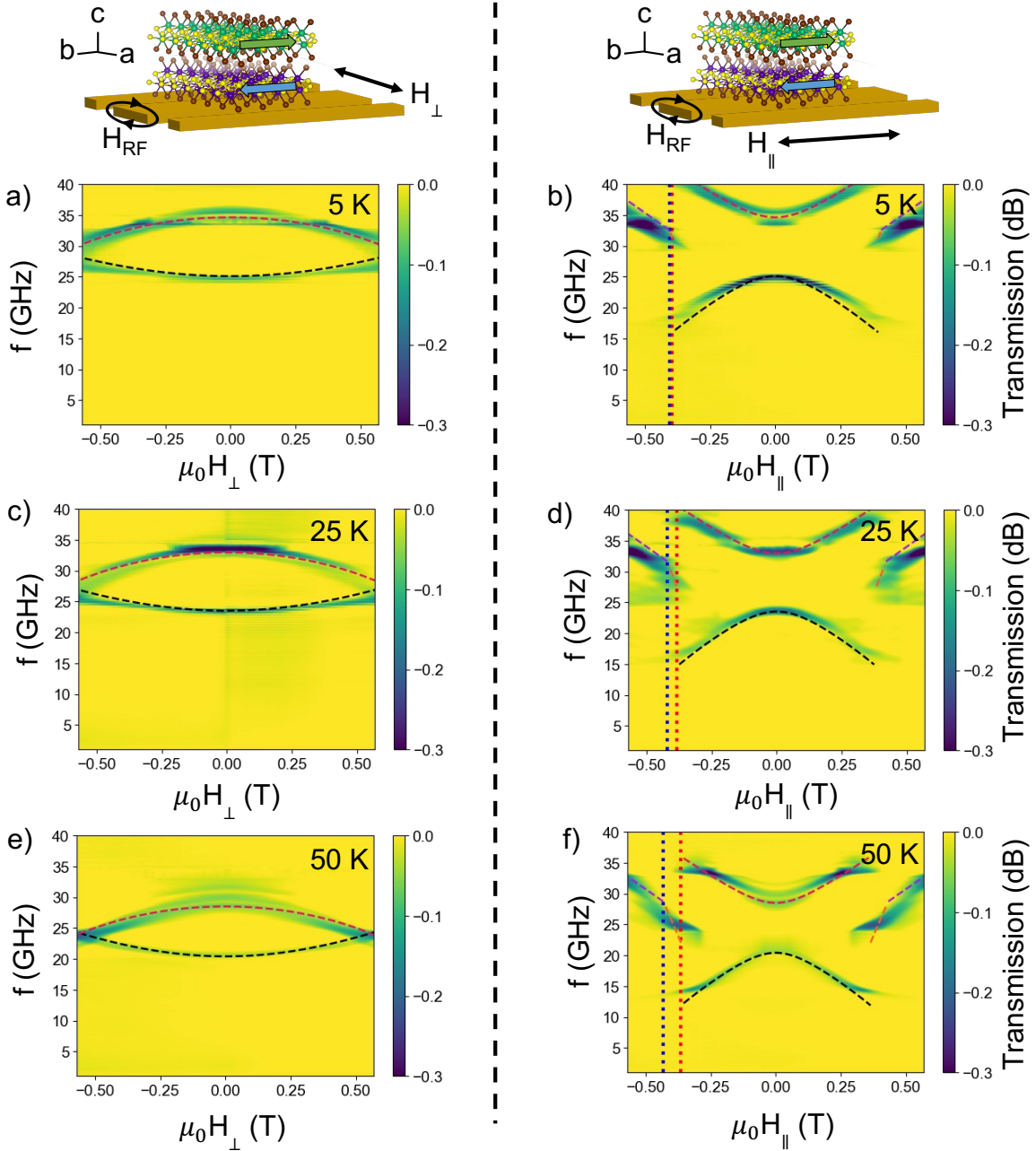

19

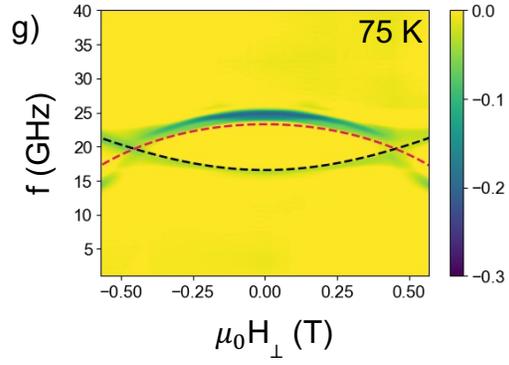
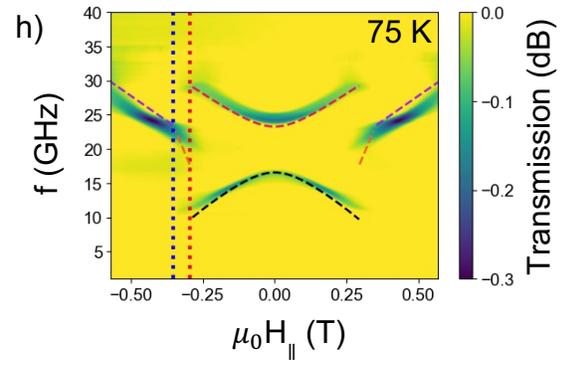
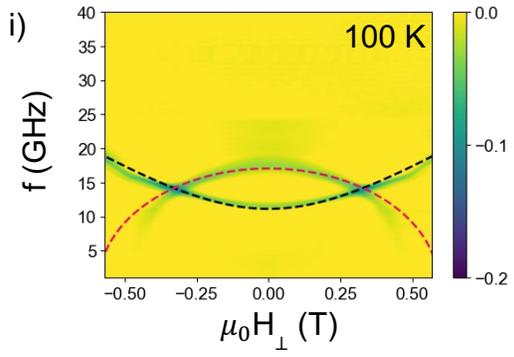
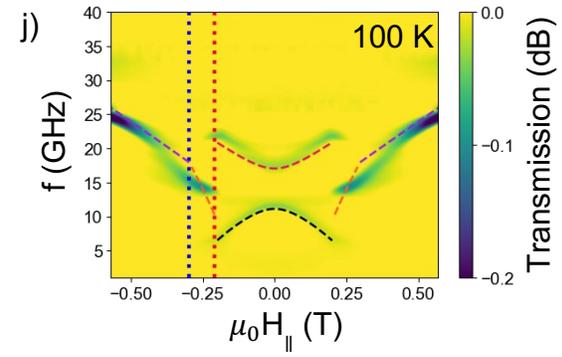
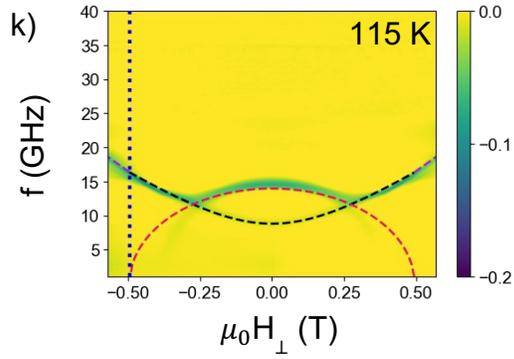
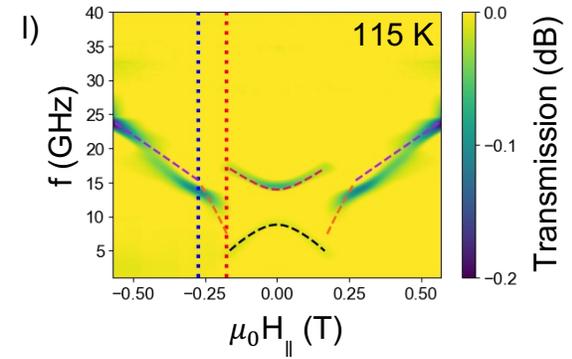
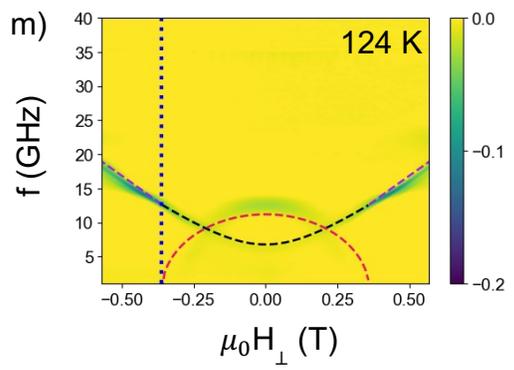
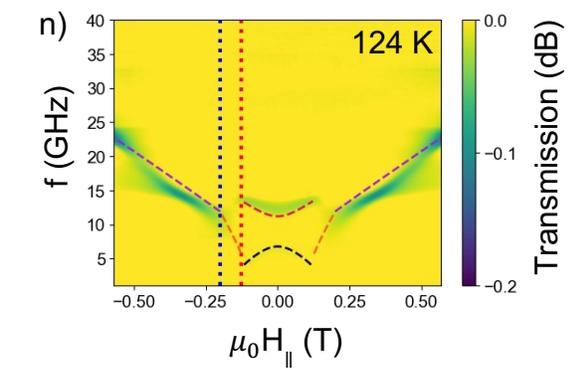

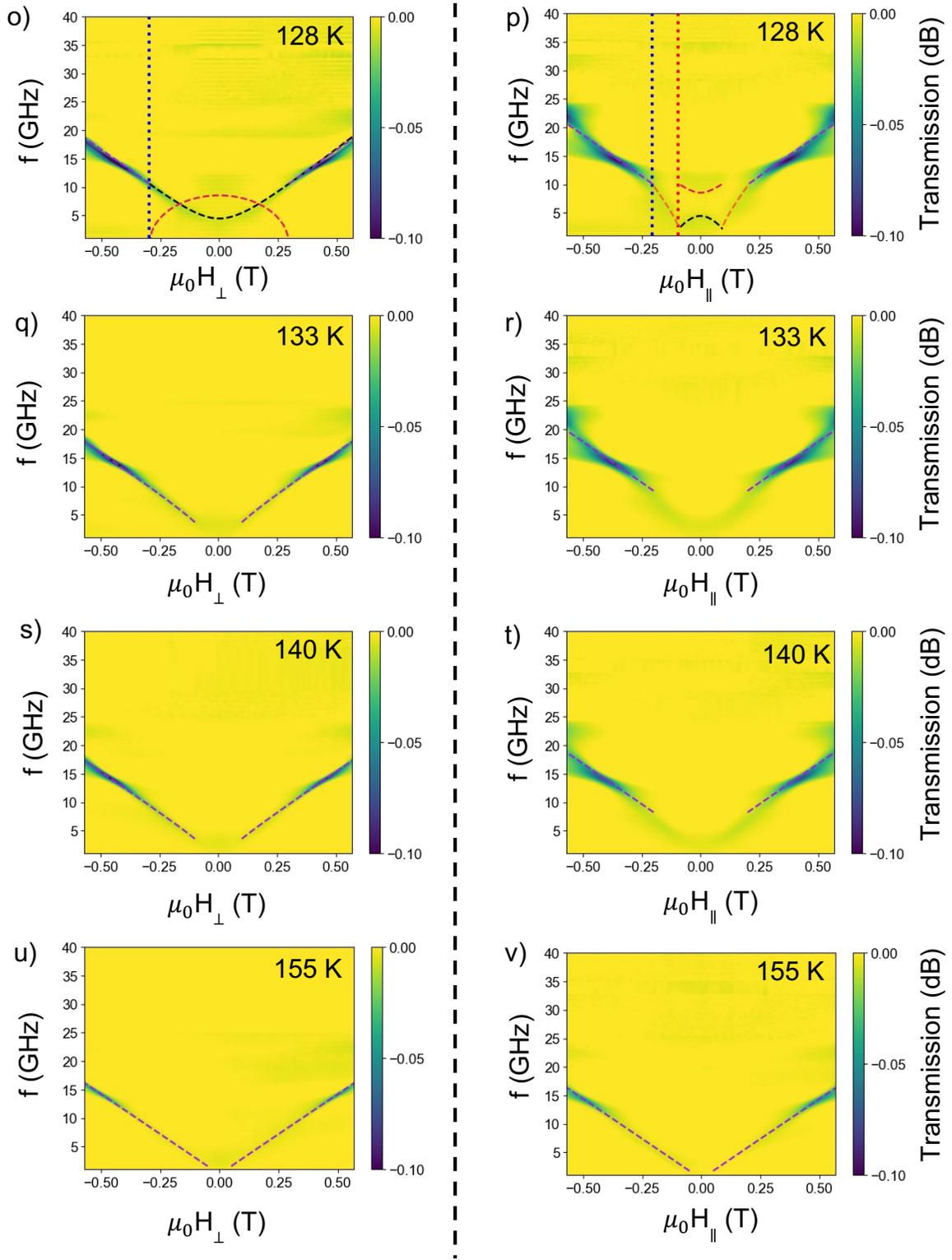

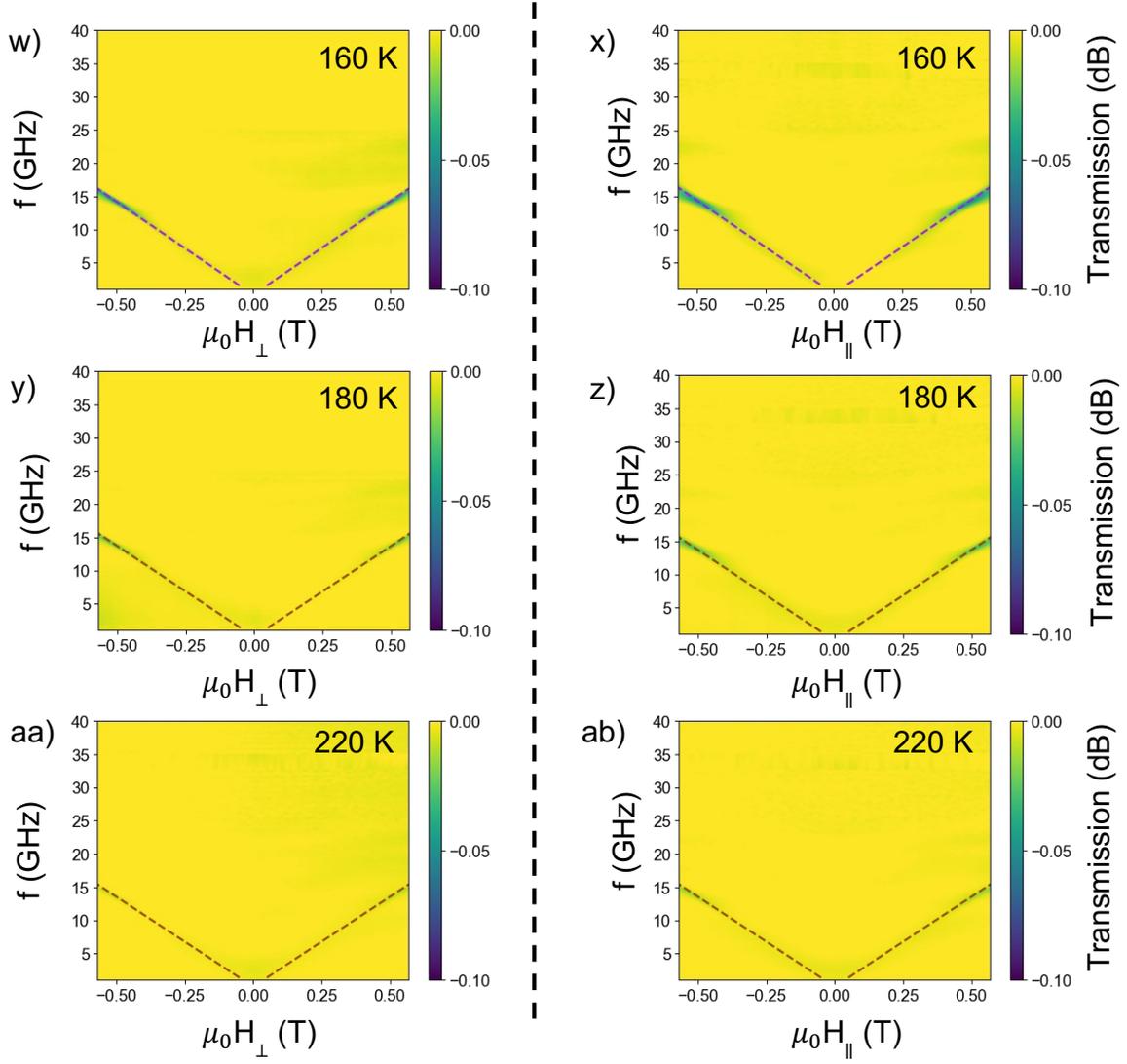

FIG. S8: Transmission spectra at all measured temperatures with fitted magnetic resonance modes. Red and blue dotted lines indicate the fitted spin-flop and ferromagnetic alignment fields respectively.

## VI. SPECTRA FOR A SECOND CRYSTAL WITH ITS a-AXIS PARALLEL TO THE COPLANAR WAVEGUIDE CENTER LINE

We have also measured the spectra of a second CrSBr crystal oriented with its long edge ($a$-axis) parallel to the center line of the coplanar waveguide (Fig. S9). In this configuration, for magnetic fields applied parallel to the easy-magnetic-anisotropy $b$ axis ($H_\parallel$) we measure similar resonances compared to those in the main text – two modes at low fields followed by a discontinuous jump leading eventually to a ferromagnetic-like mode. For applied magnetic fields aligned along the intermediate-anisotropy $a$ axis ($H_\perp$), due to 2-fold rotational symmetry in this configuration (see supplementary information, Section VII), only the acoustic mode is excited (the one odd under the symmetry). Figure S10 shows the anisotropy and exchange field strengths as a function of temperature extracted from fits to these spectra. All parameters are in reasonable agreement with those extracted for the sample analyzed in the main text.

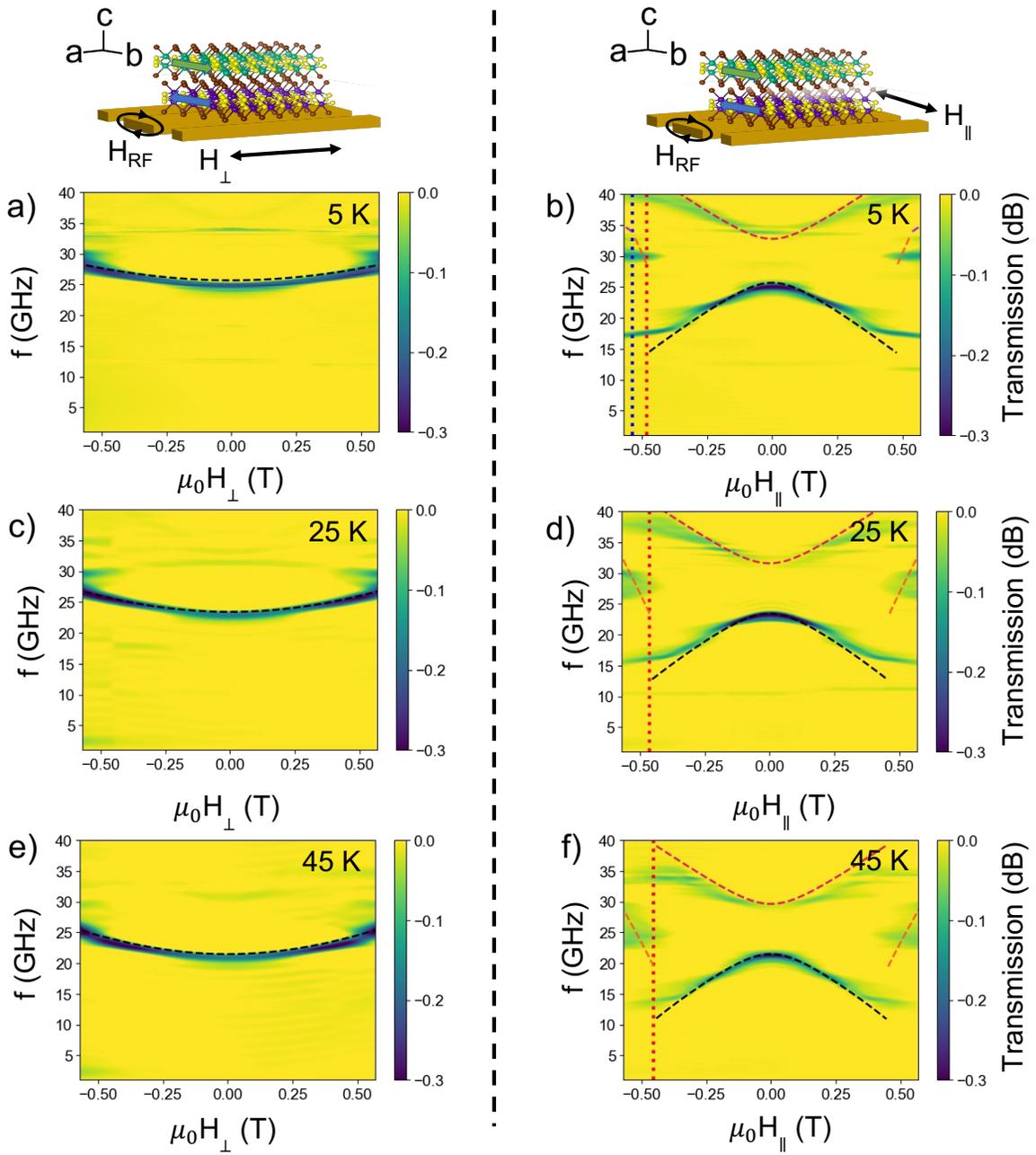

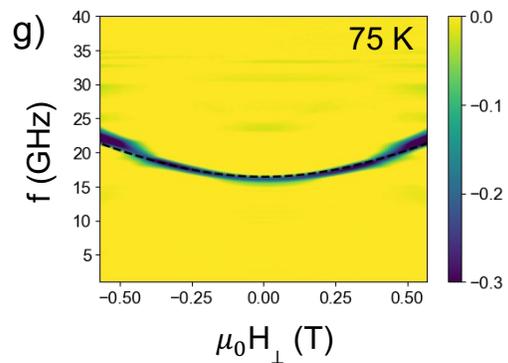
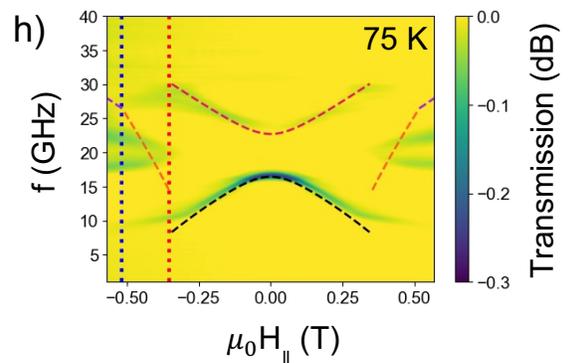
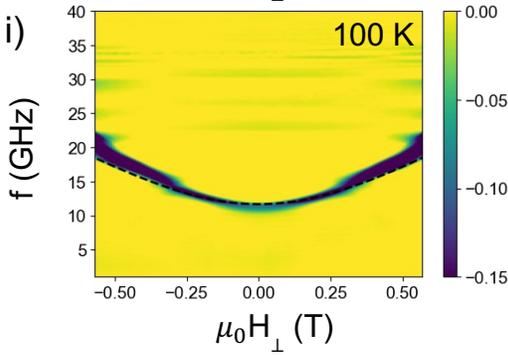
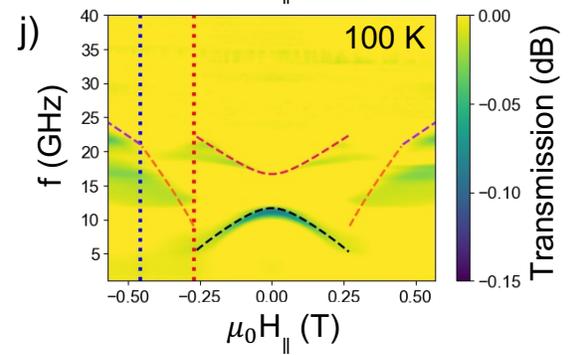
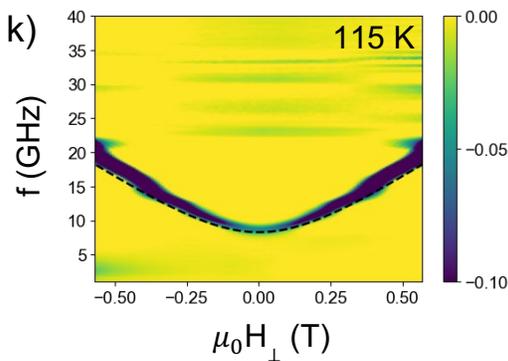
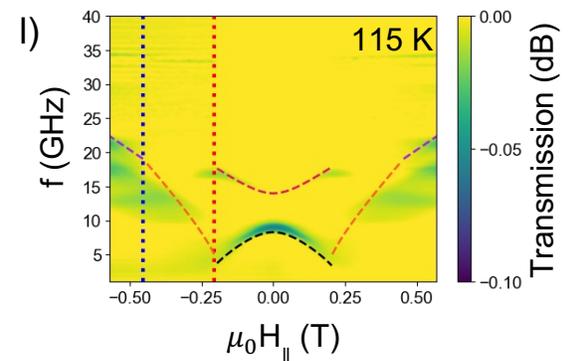
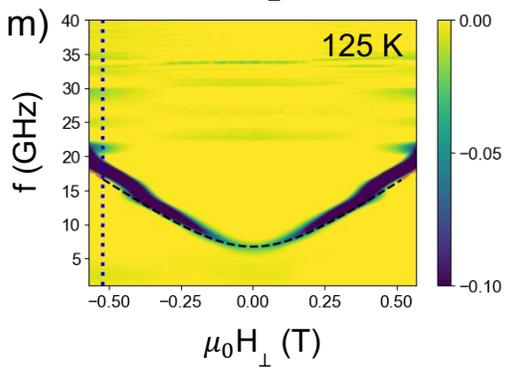
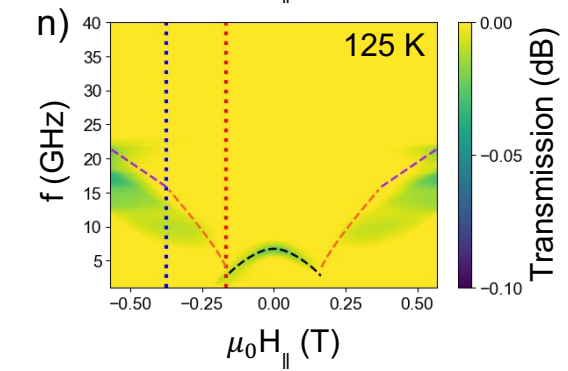

25

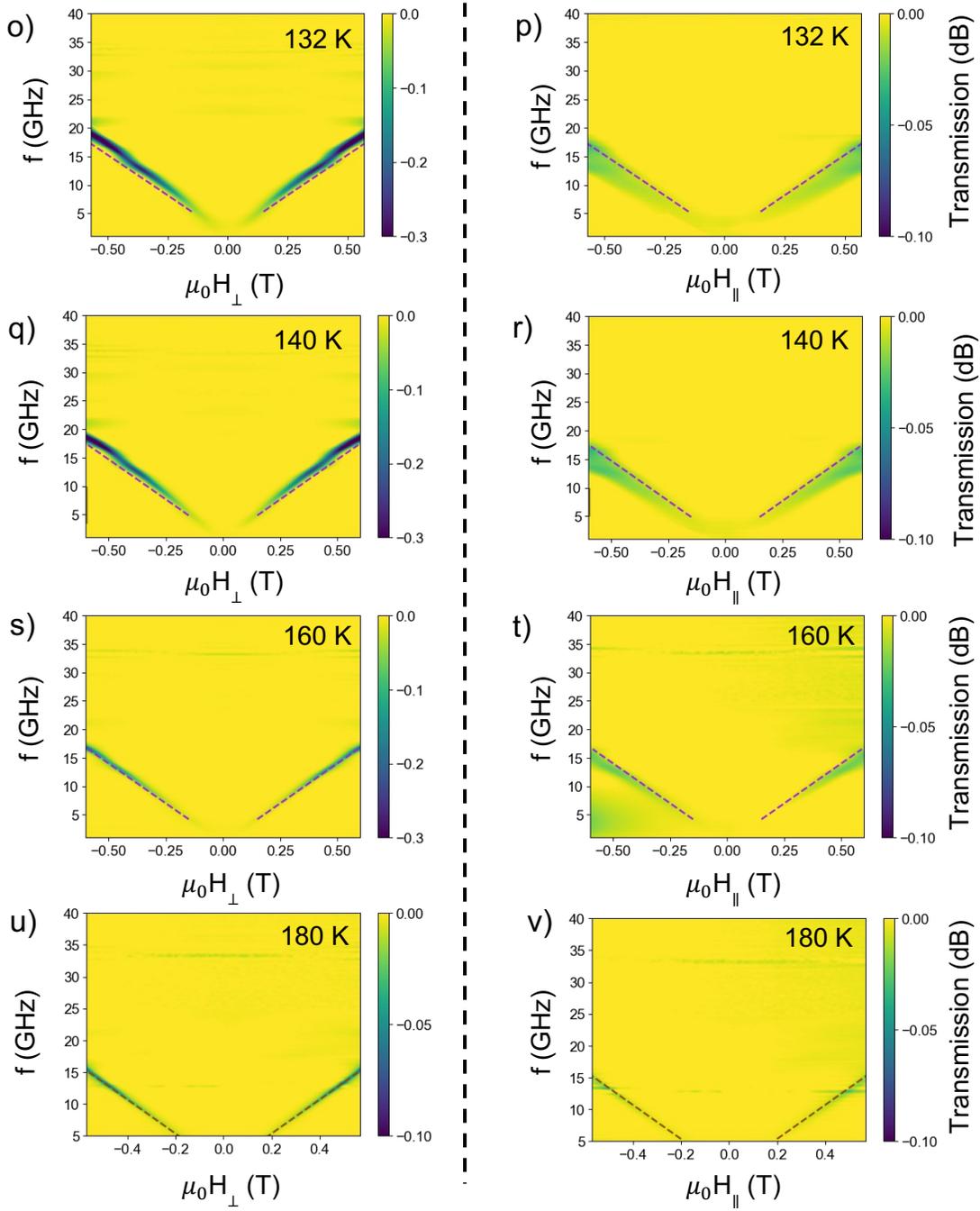

FIG. S9: Transmission spectra at different temperatures for the second crystal, with *a*-axis parallel to the center line of the coplanar waveguide.

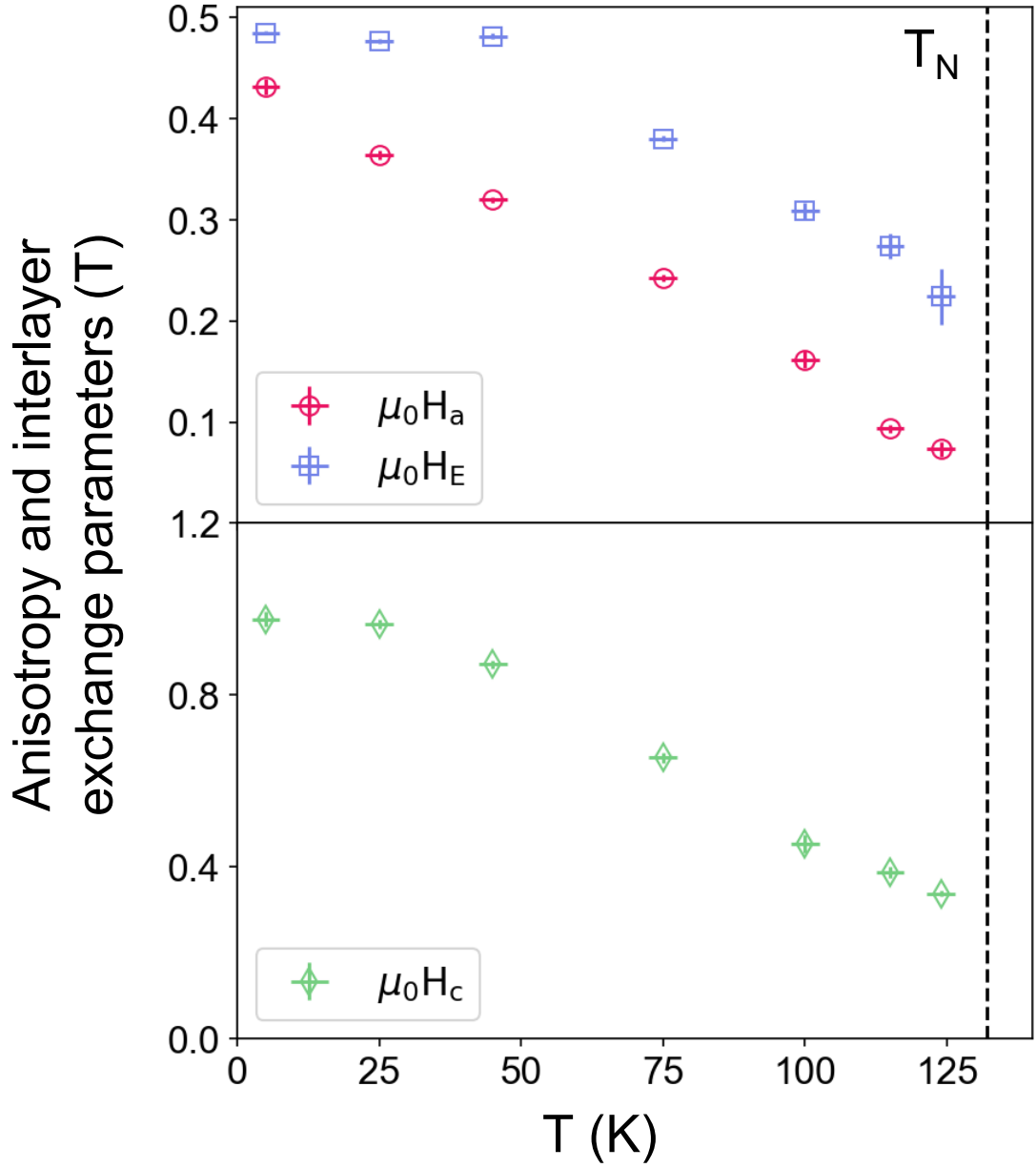

FIG. S10: Exchange and anisotropy fit parameters as a function of temperature for crystal 2, with *a*-axis ∥ to the center line of the coplanar waveguide. The dashed line indicates the $T_N$ previously reported for bulk CrSBr [6, 7].

# VII. SELECTIVE EXCITATION OF MODES FROM 2-FOLD ROTATION SYMMETRY CONSIDERATIONS

For the $H_\perp$ configuration of crystal 2 with field applied along $\hat{a}$ (SI Section VI), we observe only the acoustic modes, not the optical modes. This arises from symmetry considerations previously discussed for $CrCl_3$ [3], due to the system being symmetric under two-fold rotation about the $\hat{a}$-axis and exchange of the spin sublattices. Within same formalism used in ref. [3], the two-fold rotational symmetry still applies for a sample with triaxial anisotropy, with an intermediate anisotropy axis parallel to $\hat{a}$. Including terms to first-order in precession amplitudes, the applicable L-L equations are:

$$i\omega\delta\boldsymbol{m}_1 = \gamma\boldsymbol{m}_1^{eq} \times (H_{eq}\delta\boldsymbol{m}_1 + H_E\delta\boldsymbol{m}_2 + H_c(\delta\boldsymbol{m}_1 \cdot \hat{c})\hat{c} + H_a(\delta\boldsymbol{m}_1 \cdot \hat{a})\hat{a}) + \boldsymbol{\tau}_1$$
$$i\omega\delta\boldsymbol{m}_2 = \gamma\boldsymbol{m}_2^{eq} \times (H_{eq}\delta\boldsymbol{m}_2 + H_E\delta\boldsymbol{m}_1 + H_c(\delta\boldsymbol{m}_2 \cdot \hat{c})\hat{c} + H_a(\delta\boldsymbol{m}_2 \cdot \hat{a})\hat{a}) + \boldsymbol{\tau}_2.$$
(35)

Here, $\boldsymbol{m}_{1(2)}^{eq}$, $\delta\boldsymbol{m}_{1(2)}^{eq}$ are the equilibrium and displacement vectors of each sublattice, respectively. $\boldsymbol{\tau}_{1(2)}$ are the torques due to the RF field $\boldsymbol{H}_{RF}$ acting on each sublattice. $H_{eq}$ is the magnitude of total effective magnetic field at the equilibrium condition: $H_{eq}\boldsymbol{m}_{1(2)}^{eq} = H_\perp\hat{a} - H_E\boldsymbol{m}_{2(1)} - H_a(\boldsymbol{m}_{1(2)} \cdot \hat{a})\hat{a}$. By balancing the torques at equilibrium, $H_{eq} = |H\hat{a} - H_E\boldsymbol{m}_2 - H_a(\boldsymbol{m}_1 \cdot \hat{a})\hat{a}| = H_E$. Because the spin-flop configuration is 2-fold symmetric about $\hat{a}$, $C_{2a}\hat{\mathbf{m}}_1^{eq} = \hat{\mathbf{m}}_2^{eq}$, where $C_{2a}$ is an operator for a 2-fold rotation about the a axis, we can decouple Eqs. (35) by defining two orthogonal modes that are linear combinations of the sublattice displacement vectors: $\delta\boldsymbol{m}_\pm = \delta\boldsymbol{m}_1 \pm C_{2a}\delta\boldsymbol{m}_2$, that obey the L-L equation

$$i\omega\delta\boldsymbol{m}_\pm = \gamma m_1^{eq} \times (H_e\delta\boldsymbol{m}_\pm \pm H_e C_{2y}\delta\boldsymbol{m}_\pm + H_c(\delta\boldsymbol{m}_\pm \cdot \hat{c})\hat{c} + H_a(\delta\boldsymbol{m}_\pm \cdot \hat{a})\hat{a}) + \tau_\pm \quad (36)$$

Using the equilibrium position condition previously obtained through minimizing the free energy (Eq. (7)): $\sin\chi = H_\perp/(2H_E + H_a)$, Eq. (36) can be expanded to give a $6 \times 6$ block diagonal matrix, with upper left and lower right quadrants corresponding to the $+$ and $-$

orthogonal modes respectively

$$\begin{vmatrix} -iw & 0 & -\gamma H_c \sin\chi & 0 & 0 & 0 \\ 0 & -iw & \gamma H_c \cos\chi & 0 & 0 & 0 \\ 0 & -\gamma(2H_E + H_a)\cos\chi & -iw & 0 & 0 & 0 \\ 0 & 0 & 0 & -iw & 0 & -\gamma(2H_E + H_c)\sin\chi \\ 0 & 0 & 0 & 0 & -iw & \gamma(2H_E + H_c)\cos\chi \\ 0 & 0 & 0 & 2\gamma H_E \sin\chi & -\gamma H_a \cos\chi & -iw \end{vmatrix} = 0. \quad (37)$$

Solving for the eigenvalues, we get the same solutions $\omega_{1,\perp}$ and $\omega_{2,\perp}$ previously calculated from the coupled L-L equations (SI. IV, Eqs. (8,9)). We note that the acoustic mode is given by the − mode ($\omega_{1,\perp} = \omega_-$) where $\delta\bm{m}_1^{eq}$ and $\delta\bm{m}_2^{eq}$ rotate in-phase, while the optical mode is given by the + mode ($\omega_{2,\perp} = \omega_+$), where $\delta\bm{m}_1^{eq}$ and $\delta\bm{m}_2^{eq}$ rotate out-of-phase.

The absence of the $\omega_{2,\perp}$ optical mode in our spectra can be understood as follows. In general, we can split $\bm{H}_{RF}$ into components that are even or odd upon 2-fold rotation about $\hat{a}$. For the even components, using $\hat{\bm{m}}_1^{eq} = C_{2a}\hat{\bm{m}}_2^{eq}$ and $\bm{H}_{RF} = C_{2a}\bm{H}_{RF}$, we can deduce that the torques on sublattices 1, 2 also have even symmetry.

$$\begin{aligned} \bm{\tau}_1 &= \hat{\bm{m}}_1^{eq} \times \bm{H}_{RF} = C_{2a}\hat{\bm{m}}_2^{eq} \times C_{2a}\bm{H}_{RF} \\ &= C_{2a}(\hat{\bm{m}}_2^{eq} \times \bm{H}_{RF}) = C_{2a}\bm{\tau}_2 \end{aligned} \quad (38)$$

From the definitions of $\bm{\tau}_\pm$, this gives $\bm{\tau}_- = 0$ and $\bm{\tau}_+ \neq 0$. In other words, components of $H_{RF}$ that are even with rotation about $\hat{a}$, excite the optical mode but not the acoustic mode. A similar argument can be made for $\bm{H}_{RF}$ components that are odd with rotation about $\hat{a}$, which gives $\bm{\tau}_1 = -C_{2y}\bm{\tau}_2$. This implies that $\bm{\tau}_+ = 0$ while $\bm{\tau}_- \neq 0$ so only the acoustic mode is excited.

For the $H_\perp$ configuration of crystal 2, $\bm{H}_{RF}$ only has an in-plane component along the $\hat{b}$ axis and an out-of-plane component along the $\hat{c}$ axis, both of which are odd with rotation about $\hat{a}$. Therefore, only the acoustic mode appears in the measured spectra. In the large field regime for the $H_\parallel$ configuration, the system is 2-fold symmetric around the applied field, this time along $\hat{b}$. Here, the in-plane component of $\bm{H}_{RF}$ is even, while the out-of-plane component is odd with rotation about $\hat{a}$. Following the same arguments above, we deduce that both the optical and the lower frequency ferromagnetic modes should be excited.

## VIII. MICROWAVE POWER DEPENDENCE

Here we present examples of spectra acquired using various applied microwave powers from -40 dB to -5 dB at 100 K. As there is negligible variation in the measured resonance modes with power, we conclude that there is no significant sample heating from microwave excitation in our measurements. We use -20 dB for all measurements unless otherwise specified.

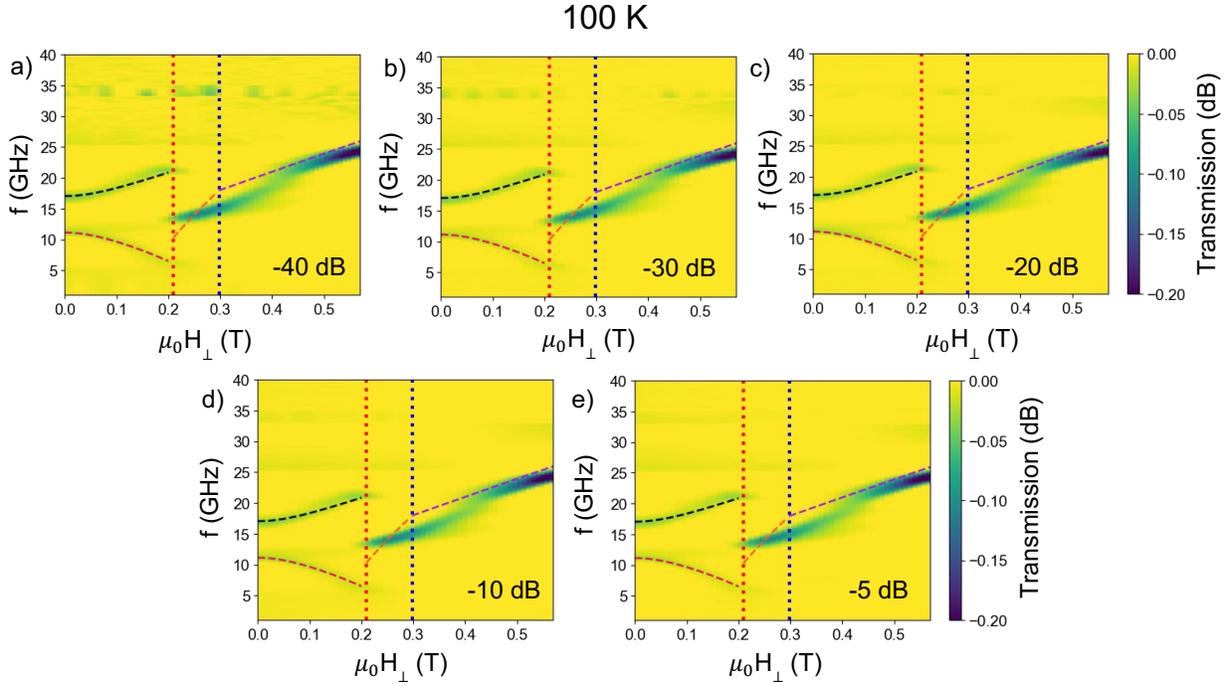

FIG. S11: Absorption spectra at various applied microwave powers from -40 dB to -5 dB at 100 K.

\clearpage

\end{document}